\documentclass[11pt,onecolumn,authoryear]{elsarticle}
\usepackage[a4paper,margin=2.5cm]{geometry}
\usepackage{setspace}
\usepackage[T1]{fontenc}
\usepackage[utf8]{inputenc}
\usepackage{lmodern}
\usepackage[english]{babel}
\usepackage{caption}

\usepackage{amsmath}
\catcode`\@=11 \@addtoreset{equation}{section}

\catcode`\@=12
\usepackage{amssymb}
\usepackage{amsbsy} 
\usepackage{amsfonts}

\usepackage{color}
\usepackage{xcolor}
\usepackage{graphics}
\usepackage{epsfig,psfrag,graphicx}
\usepackage{eepic,epic}
\usepackage{dsfont}
\usepackage{tabularx}
\usepackage{tablefootnote}
\usepackage{longtable}
\usepackage{booktabs}

\usepackage{enumerate}
\usepackage{epsfig} 

\usepackage{comment}

\usepackage{hyperref}

\usepackage{doi}

\usepackage{tablefootnote}
\usepackage{nicefrac}
\usepackage{multirow}
\usepackage{siunitx}

\usepackage{float}
\usepackage{pifont}

\usepackage{appendix}

\usepackage{vmargin}
\setmarginsrb  { 0.7in}  
               { 0.6in}  
               { 0.7in}  
               { 0.8in}  
               {  20pt}  
               {0.25in}  
               {   9pt}  
               { 0.3in}  

\newtheorem{definition}{Definition}[section]

\newtheorem{remark}[definition]{Remark}

\newcommand{\nc}{\newcommand}

\nc{\proof}[1]{{\bf Proof of Proposition {#1}: }}

\allowdisplaybreaks

\journal{Omega} 

\title{\LARGE A Two-Stage Decision Support System for Sustainability-Aware Long–Short Portfolio Optimization}

\begin{document}

\begin{frontmatter}


\author[univpm]{Giacomo di Tollo}
\ead{g.ditollo@staff.univpm.it}

\author[units]{Massimiliano Kaucic\corref{cor1}}

\author[units]{Filippo Piccotto}
\ead{filippo.piccotto@deams.units.it}

\address[univpm]{\textsc{Marche Polytechnic University}, \textit{Department of Management}, Piazza Roma 22, Ancona, Italy}

\address[units]{\textsc{University of Trieste}, \textit{Department of Economics, Business, Mathematics and Statistics}, Via Valerio 4/1, Trieste, Italy}

\cortext[cor1]{Corresponding author. E-mail address: massimiliano.kaucic@deams.units.it (M. Kaucic).}

\begin{abstract}
This paper proposes a two-stage decision support system for long-short portfolio optimization under environmental, social, and governance (ESG) considerations. In the first stage, assets are evaluated using a multi-criteria procedure based on TODIMSort, with criterion weights derived using the MEREC (Removal Effects of Criteria) method. This allows assets to be assigned to classes ordered according to preferences that respond to market conditions and investor priorities, thus generating sets of long and short opportunities that dynamically adapt to the prevailing regime. In the second stage, we formulate a non-convex portfolio optimization problem that maximizes the Omega ratio while respecting budget, bound and leverage constraints. To solve it, we introduce an adaptive particle swarm solver equipped with a controller that selects, at each iteration, the most suitable recombination operator from a diverse pool of operators and combines it with a projection-based repair mechanism for constraint management. The empirical study, conducted on 421 stocks in the STOXX Europe 600 index, examines both the exploration capabilities and solution quality of the proposed solver compared to state-of-the-art benchmarks, as well as the ex post profitability of the resulting portfolio strategies. The results show that ESG-enhanced long-short portfolios offer competitive and often superior performance compared to their non-ESG counterparts and the market-value-weighted benchmark.
\end{abstract}


\end{frontmatter}

\section{Introduction}

The mean-variance analysis (\cite{Markowitz1952}) represents the cornerstone of modern portfolio theory and is widely accepted as a wealth management tool in the financial industry (\citet{Guerard2009, Kalayci2019}). Specifically, this model illustrates the portfolio selection problem as a quadratic program that minimizes portfolio variance given a level of expected return. Beyond this foundational framework, several studies proposed alternative model formulations and incorporated various types of constraints to account for specific investment guidelines and institutional features of the investment process. On the one hand, advances in portfolio modeling include using other measures to quantify portfolio risk, such as mean absolute deviation, value-at-risk, and conditional value-at-risk (\citet{KonnoYamazaki1991, RockafellarUryasev2000, Krokhmal2002}). In addition, investors are typically interested in the reward component represented by the expected return of the portfolio, and in the performance of the proposed investment strategy compared to a market benchmark. Hence, many authors proposed optimization models in which benchmark-related reward-to-risk measures guide investors' decisions. Some examples include the Sharpe ratio and its extensions (\citet{Kaucic2024}), the Omega ratio (\citet{Guastaroba2016, Sharma2017}), and other performance metrics that incorporate tail-risk measures (\citet{Farinelli2008, Corazza2021}). On the other hand, a substantial body of literature introduced real-world portfolio constraints in addition to standard budget, no short-selling, and floor/ceiling constraints. First, cardinality constraints limit a portfolio to have a maximum number of constituents \citep{Chang2000}. Then, turnover constraints control the magnitude of trades during portfolio rebalancements (\citet{Soleimani2009, Woodside2011}), liquidity constraints impose a limit on the amount an individual can borrow (\citet{Lo2006}), or transaction costs constraints control the amount paid when buying or selling portfolio positions (\citet{Borkovec2010, RuizSuarez2015}). Finally, to foster weights diversification, some studies introduced constraints on portfolio norms (\citet{DeMiguel2009}), or included risk-budgeting constraints in portfolio design (see \citet{Kaucic2019} and references therein).

Although long-only portfolio selection simplifies optimization and is the standard practice among scholars and practitioners, it may lead to several pitfalls. In fact, forbidding short positions restricts diversification and limits potential opportunities to exploit negative views on assets and construct market-neutral strategies. \citet{Jacobs1999} proposed a long-short framework applied to the Markowitz mean-variance analysis, laying the foundations for the subsequent long-short portfolio management literature. \citet{Gilli2011} developed a portfolio model that maximizes the Omega ratio subject to cardinality constraints, which employs the 130/30 rule, where short sales are allowed up to 30$\%$ of the net asset value of the overall portfolio. However, enriching the mean-variance framework through alternative performance measures and realistic constraints significantly increases the complexity of the portfolio selection problem, often leading to mixed-integer, non-linear, and non-convex formulations, for which finding feasible and optimal solutions becomes computationally challenging using exact solvers (\citet{MoralEscudero2006}). For these reasons, metaheuristic approaches, such as evolutionary and swarm intelligence algorithms equipped with hybrid constraint-handling techniques, have been developed to optimize complex portfolio optimization problems (see \citet{ErwinEngelbrecht2023} for a comprehensive literature review). In particular, several authors used metaheuristic algorithms in long-short portfolio formulations (\citet{PaiMichel2012, Chou2021, DiTollo2024}).

In parallel, a rapidly growing stream of research recently explored the use of machine-learning methods, and in particular learn-to-rank (LTR) algorithms, for long–short portfolio construction. These approaches aim at generating ordered lists of assets from which long and short legs are derived directly. Early contributions such as \citet{Zhang2022} propose list-wise ranking architectures in which the model is trained to maximize the separation between top- and bottom-ranked assets, while more recent developments refine this idea through enhanced ranking losses and feature engineering \citep{Linger2025}. Although promising, this line of work revealed important limitations, since many LTR-based long–short strategies tend to produce unstable rankings and exhibit pronounced sensitivity to noise in the predictors, which ultimately undermines the robustness of the resulting allocations and makes performance heavily dependent on the particular choice of
hyperparameters and training windows. To address these issues, \citet{Kouloumpris2024} introduced SABER, a stochastic-aware bootstrap ensemble ranking method designed to improve stability and mitigate overfitting by aggregating multiple perturbed ranking estimates. While SABER enhances robustness relative to single-model LTR methods, it still operates as a purely data-driven ranking engine and does not explicitly incorporate behavioral or financial-interpretation principles in the formation of long and short sets.

To overcome these limitations, in this paper we propose an integrated two-stage decision support system (DSS) to construct equity portfolios that also allows for short selling. We consider the asset allocation problem from the perspective of institutional investors and fund managers operating in equity markets, who base their financial decisions on the Omega ratio performance measure and incorporate ESG considerations in their decision process. The central premise of our contribution is that the definition of long and short legs should arise from a data-driven and behaviorally coherent decision process rather than from ad-hoc statistical ranking procedures. 

In the proposed architecture, stage 1 performs stock screening using multi-criteria decision-making (MCDM) theory. This module starts by identifying the current market phase, distinguishing among bull, bear, and sideways regime. Then, it integrates phase-independent and phase-dependent criteria in the stock-picking process.
To objectively determine the relative importance of the considered criteria, we employ the MEREC method (\citet{Ghorabaee2021}). Then, we obtain assets' rankings through the so-called TODIM (the portuguese acronym for interactive multi-criteria decision making), a MCDM method introduced by \citet{GomesLima1991} that integrates behavioral finance principles to evaluate criteria alternatives. This method has been successfully applied in the financial field, due to its capacity to reflect behavioral characteristics of decision-makers (\citet{AlaliTolga2019, Wu2022}). However, to construct long-short portfolios, a ranking is not sufficient, since we need to assign assets to three classes, namely long, neutral, and short. To achieve this goal, we utilize the TODIMSort (\citet{Wang2023}), an enhanced version of the TODIM method that assigns assets to ordered preference classes through limiting profiles. A key novelty of our approach is that the limiting profiles are not specified solely on the basis of subjective preferences, as in traditional TODIMSort implementations, but are instead constructed dynamically by combining investor-specified percentile thresholds with empirical distributions of criteria conditioned on the prevailing market regime. As a result, limiting profiles are both behaviorally meaningful and consistent with the prevailing market regime. Once the investment set is defined, stage 2 consists of a non-convex portfolio optimization model that maximizes the Omega ratio under long–short budget and bound constraints. To solve this problem, we develop an adaptive multi-operator particle swarm optimization algorithm (AMPSO), which combines time-varying acceleration coefficients \citep{Ratnaweera2004}, an adaptive operator selection mechanism \citep{filograsso2023}, and a projection-based constraint-handling procedure to ensure feasibility \citep{maculan2003}. The Supplementary Material contains a graphical chart that illustrates our two-stage DSS architecture.

To evaluate the proposed framework, we conduct a threefold extensive empirical analysis employing a large panel of European equities over a 10–year horizon, integrating daily prices, monthly ESG indicators, and regulatory constraints on short selling. First, we examine the behavior of the MEREC–TODIMSort module, analyzing how market phases influence criteria weights, the shape of limiting profiles, and the resulting long, neutral,
and short sets. Second, we evaluate the optimization stage through a benchmark suite of long–short portfolio
instances, comparing AMPSO with multiple PSO-based variants under identical computational budgets. Finally, we assess the out-of-sample performance of the resulting strategies under different leverage levels and limiting-profile configurations.  Additional portfolios that exclude ESG-based criteria are constructed to isolate the marginal contribution of sustainability-related information to overall performance.

The main contributions of this study can be summarized as follows.
\begin{itemize}
    \setlength{\itemsep}{0pt}
    \item We design a unified architecture for long-short portfolios in which a multi-criteria screening procedure and a non-convex optimization model are structurally interconnected, with ESG indicators directly embedded in the pre-selection stage.
    \item We introduce a novel implementation of TODIMSort in which limiting profiles are constructed dynamically using investor-defined percentile thresholds combined with empirical criterion distributions conditioned on the prevailing market phase. This yields a data-driven, market-coherent and behaviorally consistent segmentation of assets into short, neutral and long classes.
    \item We develop a variant of the particle swarm optimization algorithm that integrates time-varying acceleration coefficients, adaptive operator selection, and  projection repair mechanisms.
\end{itemize}
\noindent The remainder of the paper is organized as follows. 
Section~\ref{section:literature} positions our contribution within the existing literature on multi-criteria decision support systems and long–short portfolio optimization. Section~\ref{section:PortfolioModel} introduces the proposed long–short portfolio formulation based on the Omega ratio and discusses its financial interpretation. Sections~\ref{section:MCDM} and \ref{section:OptimizationAlgorithm} describe the MEREC-TODIMSort screening framework and the AMPSO algorithm, respectively. Section~\ref{section:experiments} reports the empirical analysis, covering data specification, screening behavior, algorithmic evaluation, and ex-post profitability. Finally, Section~\ref{section:conclusions} concludes and indicates directions for future research.

\section{Literature Positioning}\label{section:literature}
Table \ref{tab:literature} reports some recent selected studies that propose two-stage decision support systems for constructing portfolio strategies. We selected the papers that respected two conditions: (i) the proposal of a DSS in the field of portfolio optimization, and (ii) the presence of a long-short portfolio formulation, and/or the presence of ESG considerations. Only four papers in the recent literature satisfied these requirements. As an exception, we report the work of \citet{Xidonas2021}, since it represents a seminal contribution in this field. In this paper, the authors employ four MCDM methods, namely ELECTRE III, PROMETHEE II, MAUT, and TOPSIS, to perform a pre-selection of 60 stocks from a subset of 516 NYSE-listed assets. Then, they introduce a long-only multi-objective portfolio model that maximizes the expected return while minimizing the portfolio risk, also accounting for many realistic constraints, and solve the resulting problem using a Genetic Algorithm (GA). \citet{Banerjee2024} propose a two-stage DSS where a fuzzy extension of TODIM performs the stock selection. In addition, they consider a mean–variance model with cardinality, round lots, transaction costs, and budget constraints that also allows for short sales. They employ a GA algorithm to find optimal solutions and assess the capabilities of this model across three major equity markets, namely Nifty 100 (India), S\&P 500 (USA), and FTSE 100 (UK). The studies of \citet{Jain2025a, Jain2025b} propose a two-stage architecture to construct optimal portfolios with an ESG mandate using data from the Nifty 50 and Nifty 100 indices, respectively. In these works, ESG considerations enter the portfolio construction in the constraints set, as the authors include in the model a minimum threshold for the ESG score of the portfolio. Moreover, their objective function is a weighted function of higher-order moments of the return distribution. For the optimization phase, both studies employ a tailored version of the teaching-learning-based optimization algorithm (TLBO), with a repair constraint-handling strategy for bound constraints. However, they do not allow for short-sales in their models. Finally, in the work of \citet{VanWallendael2026}, the authors first define ESG-informed returns as a linear combination of asset returns and ESG scores, then employ these returns in portfolio selection by proposing a long-only mean-CVaR formulation. In addition, they also use ESG as a screening tool, eliminating assets under a minimum ESG acceptance threshold in the pre-selection phase.

\begin{table}[h!]
\centering
\scriptsize
\begin{tabularx}{\textwidth}{l X X X X c c}
\toprule
\textbf{Reference} &
\textbf{MCDM} &
\textbf{Opt. Model} &
\textbf{Constraints} &
\textbf{Solver+CH} & \textbf{ESG-Aware} & \textbf{Long-Short} \\
\midrule

\cite{Xidonas2021} &
4 MCDM methods + Entropy Weighting &
Mean-Variance MOP &
Cardinality, budget, no short-selling, bounds &
Genetic Algorithm (GA) - no CH reported & \ding{55} & \ding{55}\\

\midrule

\cite{Banerjee2024} &
Fuzzy TODIM & Mean-Variance MOP &
Cardinality, Budget, Round lot, Bounds, Transaction costs &
Genetic Algorithm (GA) - no CH reported & \ding{55} & \checkmark \\

\midrule

\citet{Jain2025a} &
VIKOR + Entropy Weighting &
Weighted function of skewness, kurtosis, Gini index &
Minimum expected return, Budget, no short-selling, Bounds, Minimum ESG &
Teaching--Learning-Based Optimization (TLBO) + repair CH & \checkmark & \ding{55} \\

\midrule

\citet{Jain2025b} &
TODIM + Entropy Weighting &
Weighted function of skewness, kurtosis, Gini index, conditional value at risk, semi-variance &
Minimum expected return, Budget, no short-selling, Bounds, Minimum ESG  &
Teaching-Learning-Based Optimization (TLBO) + repair CH & \checkmark & \ding{55} \\

\midrule

\citet{VanWallendael2026} &
Machine learning-based stock selection &
Minimum CVaR model &
Expected return, Budget, no short-selling &
Not reported & \checkmark & \ding{55} \\

\midrule

This paper & TODIMSort + MEREC weighting &
Omega ratio maximization &
Budget, long-short bounds &
Particle swarm optimizer (PSO) + adaptive operator selection + projection CH & \checkmark & \checkmark\\

\bottomrule
\end{tabularx}

\caption{Selected literature on two-stage decision support systems in portfolio selection with multi-criteria decision making screening. ``CH'' stands for constraint-handling.}
\label{tab:literature}
\end{table}

This review highlights several open questions in the existing literature. Our study addresses these issues along the following dimensions.

\begin{itemize}
\setlength{\itemsep}{0pt}
    \item Although the introduction of two-stage decision support systems for portfolio construction is a topic of growing interest in the current literature, the application of this architecture is at an early stage of development, and no existing studies have proposed DSS in a long-short setting also accounting for ESG considerations.
    \item Analyzed studies use ESG information either in the definition of portfolio returns or to impose minimum ESG requirements within the optimization model. Our contribution embeds ESG indicators at a decision-making level, using them to obtain explicit criteria for the pre-selection stage.
    \item The proposed optimization models usually involve standard mean-variance or CVaR formulations. Our work considers the Omega ratio as the objective function, introducing non-convexity component in the optimization problem.
    \item While four out of the five works employ metaheuristic solvers to find optimal solutions of the proposed models, they devote limited attention to analyzing their convergence behavior, feasibility attainment, and constraint-handling effectiveness. In this study, we explicitly assess the performance of our proposed AMPSO, also introducing comparisons with other state-of-the art solvers.
\end{itemize}

Summing up, this work fits the intersection of (i) MCDM methods for asset screening, (ii) non-convex portfolio optimization, and (iii) advanced metaheuristic design, using these tools for constructing ESG-aware long-short portfolios.

\section{Portfolio Optimization Framework}\label{section:PortfolioModel}

In this study, we consider an investment framework composed of a financial market with $N$ risky assets and define a portfolio as a $N$-dimensional vector of asset weights $w \in \mathbb{R}^N$. We assume a single-period investment horizon, meaning that the investor allocates wealth at an initial time and evaluates performance at a subsequent time corresponding to the chosen horizon. Within this mono-periodic setting, let $(\Xi,\mathfrak{F},P)$ be the probability space on which the random variables are defined and let $R = (R_1,\ldots,R_N)^\top$ be the random vector of asset returns. Hence, $R_w = w^\top R$ represents the portfolio return. Similarly, we denote by $R_b$ the stochastic return of a benchmark index. Moreover, investor decisions are guided by the Omega ratio, a performance measure introduced in \citet{KeatingShadwick2002}. Formally, it is defined as 
\begin{equation}\label{eq:omega}
\Omega(R_w; R_b) = \frac{\mathbb{E}\left ( ( R_w - R_b )_{+} \right )}{\mathbb{E}\left ( ( R_b - R_w )_{+} \right )}
\end{equation}
where $(z)_{+} = \max\{ z,0\}$, with $z \in \mathbb{R}$.
Intuitively, the numerator captures the expected gains, while the denominator quantifies the expected losses with respect to the benchmark. The Omega ratio satisfies the following desirable axiomatic properties for performance measures, as discussed in \citet{Rachev2008}:

\begin{itemize}
\setlength{\itemsep}{0pt}
    \item \emph{Monotonicity}: if a portfolio \( w \) stochastically dominates another portfolio \( w' \), then \( \Omega(R_w; R_b) \geq \Omega(R_{w'}; R_b) \);
    \item \emph{Quasi-concavity}: the upper level sets \( \{w \in \mathbb{R}^N : \Omega(R_w; R_b) \geq \alpha\} \) are convex;
    \item \emph{Scale-invariance}: \( \Omega( \lambda R_w; R_b) = \Omega(R_w; R_b) \) for all \( \lambda > 0 \);
    \item \emph{Translation-invariance}: shifting all returns by a constant affects the ratio in a predictable way.
\end{itemize}

The fact that the Omega ratio satisfies these properties makes it particularly suitable for being optimized in portfolio selection models, especially in a non-Gaussian framework that considers higher-order moments and tail behavior. 

\subsection{Problem Statement}

We consider a portfolio selection setting in which short-selling is allowed. 
Let $I_{\text{long}}$ and $I_{\text{short}}$ denote the sets of assets assigned, respectively, to the long and short legs of the strategy, with cardinalities 
$n_{\text{long}}$ and $n_{\text{short}}$, where 
$n_{\text{long}} + n_{\text{short}} = n \leq N$.
These sets are determined endogenously through the classification procedure described in Section~\ref{section:MCDM}, and therefore reflect both investor preferences and market conditions. Then, the long–short portfolio optimization problem is formulated as
\begin{equation}
\label{eq:omega_opt_long_short}
\begin{aligned}
\max_{w \in \mathbb{R}^n} \quad 
    & \Omega(R_w;\,R_b) \\
\text{s.t.} \quad 
    & \sum_{i \in I_{\text{long}}} w_i = 1 + s \\
    & \sum_{i \in I_{\text{short}}} w_i = -s \\
    & 0 \le w_i \le w_i^{u}, \quad i \in I_{\text{long}} \\
    & w_i^{\ell} \le w_i \le 0, \quad i \in I_{\text{short}}
\end{aligned}
\end{equation}
where $s\in[0,1]$ is a parameter that controls the intensity of short exposure, $w_i^{u}>0$, and $w_i^{\ell}<0$.

The first two constraints ensure that the net investment is equal to one, with a long exposure of $1+s$ financed by a short exposure of $s$. In particular, when $s=0$ and $I_{\text{short}} = \emptyset$, the model reduces to a standard long-only allocation with $\sum_{i \in I_{\text{long}}} w_i = 1$. 
For $s>0$, the short leg generates income of size $s$, which finance an additional long exposure of the same size. Thus, the total long capital employed becomes $1+s$, while maintaining a net investment of one. Higher values of $s$ correspond to more aggressive hedge-fund-like strategies, increasing directional exposure and effective leverage. Upper and lower bounds $w_i^{u}$, $w_i^{\ell}$ impose realistic position limits on long and short weights. 

When short-selling is temporarily suspended (for example, during periods of market stress or regulatory intervention) the admissible portfolio reduces to the following long-only allocation:
\begin{equation}
\label{eq:omega_opt_long_only}
\begin{aligned}
    \max_{w \in \mathbb{R}^{n_\text{long-only}}} \quad & \Omega(R_w; R_b) \\
    \text{s.t.} \quad & \sum_{i \in I_\text{long-only}} w_i = 1 \\
    & 0 \leq w_i \leq w_i^u, \quad i \in I_\text{long-only}
\end{aligned}
\end{equation}
where $I_\text{long-only}$ is the set of indices that correspond to the portfolio constituents in the long-only setting. To ensure comparable net exposure between the long–short model  \eqref{eq:omega_opt_long_short} and the long-only regime, the admissible number of long-only assets is adjusted as $ 
n_\text{long-only} = \left\lfloor\dfrac{n_\text{long}}{1+s}\right\rfloor$
where $n_{\text{long}}$ is the cardinality of the long set identified in the classification stage. The constituents of $I_{\text{long-only}}$ are selected based on the TODIM ranking, thereby maintaining consistency with the pre-screening procedure. This specification reflects the fact that, in the absence of short-selling, 
the long leg cannot be expanded through internal financing. 
The adjustment in $n_{\text{long-only}}$ ensures that the effective risk concentration remains comparable to the leveraged long–short configuration with parameter $s$.

\subsubsection{Empirical Estimation of the Omega Ratio}

In practical applications, the true distribution of returns is unknown. Consequently, we estimate the Omega ratio using historical observations of asset returns \citep{Guastaroba2016}. Given an in-sample window of $T$ observed trading periods, we denote by $r_{j,t}$ the realized return of asset $j$ observed in period $t$, with $t =1,\dots,T$. Thus, $r_t = (r_{1,t},\dots,r_{n,t})^\top$ is the return vector for period $t$, and the corresponding portfolio return is computed as $r_{w,t} = w^\top r_t$. Let $r_{b,t}$ be the realized return of the benchmark, the estimation of the Omega ratio is given by:
\begin{equation}\label{eq:orestimate}
    \widehat{\Omega}_w = \frac{\sum_{t=1}^T \max\{r_{w,t} - r_{b,t}, 0\}}{\sum_{t=1}^T \max\{r_{b,t} - r_{w,t}, 0\}}.
\end{equation}
This formulation corresponds to a discrete approximation of the metric given in Eqn. \eqref{eq:omega}, where the numerator aggregates the gains and the denominator aggregates the losses. If the denominator is zero (a situation that occurs when all observed portfolio returns exceed the benchmark threshold), the ratio value is set to $+\infty$. Notice that the empirical Omega ratio is non-convex and non-smooth in $w$, which motivates the use of derivative-free optimization methods such as the swarm-based algorithm we propose. 

\section{Stock Screening Framework} \label{section:MCDM}

The proposed procedure begins by identifying the securities in the reference market that will represent the investable universe. The process is structured into three components:
\begin{enumerate}[(i)]
\item the MEREC method, which determines the relative importance of screening criteria;
\item the TODIM asset ranking, which produces a dominance-based ordering of alternatives for long-only strategies;
\item the TODIMSort classifier, which assigns assets to preference-ordered categories for long–short strategies.
\end{enumerate}
These components jointly define the MEREC–TODIMSort architecture. Let $N$ denote the number of candidate assets and $m$ the number of screening criteria. Define the decision matrix $X = (x_{ij}) \in \mathbb{R}^{N \times m}$, where each row corresponds to an asset and each column corresponds to a criterion. Criteria are classified either as benefits, for which higher values indicate better performance, or as costs, for which lower values are preferable.

\subsection{MEREC-Based Weighting of Screening Criteria}\label{S:MEREC}

The MEREC weighting scheme quantifies the relative importance of the screening criteria. The computational steps of the procedure are summarized below.

\begin{itemize}
\setlength{\itemsep}{0pt}
    \item \textit{Positivity adjustment}. Financial data may include zero or negative values, which are incompatible with the logarithmic ratios used in MEREC. Consequently, the decision matrix $X$ must contain strictly positive entries. To enforce positivity, we propose for each criterion $j$ to compute its minimum value $x_j^{\min} = \min_{i} x_{ij}$ and apply 
    \begin{equation}
    x_{ij} \leftarrow x_{ij} - x_j^{\min} + 1, \quad i = 1,\dots,N.
    \end{equation}
    \item \textit{Normalization}. The translated matrix is normalized into $Z = (z_{ij})$ as
    \begin{equation}
        z_{ij} =
        \begin{cases}
            \dfrac{x_{ij}}{\max_{i} x_{ij}}, & \text{if $j$ is a benefit criterion},\\[8pt]
            \dfrac{\min_{i} x_{ij}}{x_{ij}}, & \text{if $j$ is a cost criterion}.
        \end{cases}
    \end{equation}
    \item \textit{Aggregated performance}. For each alternative $i$, we compute
    \begin{equation}
        S_i = \ln\!\left( 1 + \frac{1}{m} \sum_{j=1}^{m} \big|\ln(z_{ij})\big| \right).
    \end{equation}   
    \item \textit{Performance after criterion removal}. For each criterion $j$, the performance of asset $i$ after removing criterion $j$ is defined as
    \begin{equation}
        S^{(-j)}_{i} = \ln\!\Bigg( 1 + \frac{1}{m-1} \sum_{\substack{h=1 \\ h \neq j}}^{m} \big|\ln(z_{ih})\big| \Bigg).
    \end{equation}    
    \item \textit{Removal effect and weights}. The removal effect of criterion $j$ is
    \begin{equation}
        E_j = \sum_{i=1}^{N} \big| S^{(-j)}_{i} - S_i \big|.
    \end{equation}
    Finally, we normalize to obtain the MEREC weights
    \begin{equation}
        \pi_j = \frac{E_j}{\sum_{h=1}^{m} E_h}, \qquad j = 1,\dots,m.
    \end{equation}   
\end{itemize}
Criteria whose removal causes larger changes in aggregated performance receive higher weights, reflecting their stronger discriminating capability.

\subsection{Ranking Assets for Stock Screening via TODIM}\label{subsection:TODIM}

In our context, the TODIM method is used to establish a dominance-based ranking of $N$ assets according to $m$ criteria. The procedure consists of the following steps:

\begin{itemize}
\setlength{\itemsep}{0pt}
    \item \textit{Normalization}. The decision matrix $X$ is normalized on a $[0,1]$ scale as:
    \begin{equation}
        z'_{i,j} =
        \begin{cases}
            \dfrac{x_{ij} - \min_{i} x_{ij}}{\max_{i} x_{ij} - \min_{i} x_{ij}}, & \text{if $j$ is a benefit criterion},\\[6pt]
            \dfrac{\max_{i} x_{ij} - x_{ij}}{\max_{i} x_{ij} - \min_{i} x_{i,j}}, & \text{if $j$ is a cost criterion}.
        \end{cases}
    \end{equation}
    \item \textit{Pairwise dominance}. For each pair of assets $(i,k)$ and criterion $j$, compute the comparison score:
    \begin{equation}\label{eq:CS}
        CS_j(i,k) =
        \begin{cases}
            \pi_j \,(z'_{ij} - z'_{k,j})^{\eta_1}, & \text{if } z'_{ij} \ge z'_{kj},\\[6pt]
            - \xi \,\pi_j \,(z'_{kj} - z'_{ij})^{\eta_2}, & \text{if } z'_{ij} < z'_{kj}
        \end{cases}
    \end{equation}
    where $\pi_j$ are the criterion weights obtained from MEREC, $\eta_1,\eta_2 > 0$ are sensitivity parameters for gains and losses, and $\xi > 1$ models loss aversion.
    \item \textit{Global dominance score}. For each asset $i$, aggregate the pairwise dominance contributions across all criteria and opponent assets:
    \begin{equation}
        GS_{\mathrm{raw}}(i) = \sum_{k=1}^{N} \sum_{j=1}^{m} CS_j(i,k)
    \end{equation}
    where the self-comparison term $k=i$ is null by construction. Finally, apply a min--max normalization to obtain scores in $[0,1]$:
    \begin{equation}
        GS(i) = \frac{GS_{\mathrm{raw}}(i) - \min_{\ell} GS_{\mathrm{raw}}(\ell)}
                     {\max_{\ell} GS_{\mathrm{raw}}(\ell) - \min_{\ell} GS_{\mathrm{raw}}(\ell)}.
    \end{equation}
\end{itemize}

The resulting scores enable a psychologically consistent ranking of assets, integrating performance and behavioral considerations into the selection process. This step provides a robust pre-screening phase that precedes the portfolio optimization problem \eqref{eq:omega_opt_long_only}.

\subsection{Classifying Assets for Stock Screening via TODIMSort}
\label{subsection:TODIMSort}

Although the ranking produced by the TODIM method is well suited for long-only stock selection \citep{Kaucic2025}, the construction of long-short portfolios requires a classification scheme that assigns each asset to one of several categories according to ordered preference. To this end, we employ TODIMSort \citep{Wang2023}, an extension of TODIM designed to classify alternatives into predefined classes using limiting profiles as reference points.

\paragraph{Class structure} Let $\mathcal{C}=\{\text{short},\text{neutral},\text{long}\}$
denote the ordered set of classes. Following the standard TODIMSort framework, the boundaries between adjacent classes are determined by the two limiting profiles 
$LP_1=(lp_{1,1},\ldots,lp_{1,m})$ and $LP_2=(lp_{2,1},\ldots,lp_{2,m}),$
 where $LP_1$ separates the short and neutral classes and $LP_2$ separates the neutral and long classes. Each limiting profile represents a hypothetical alternative that defines the minimum performance required to enter the corresponding higher-ranked class.

\paragraph{Dynamic, data-driven limiting profiles}
A key novelty of our approach is that the limiting profiles are not specified solely on the basis of subjective preferences, as in traditional TODIMSort implementations, but are instead constructed by combining investor preferences with empirical information extracted from historical data. This yields limiting profiles that are both behaviorally meaningful and consistent with the prevailing market regime.

Given the market phase $\phi \in \{\text{bull},\text{bear},\text{sideways}\}$ 
prevailing at the evaluation time, we identify all historical evaluation dates whose market 
phase coincides with $\phi$. 
For each such date $t'$, let 
$A^{(t')} = (x_{ij}^{(t')}) \in \mathbb{R}^{N \times m}$ 
denote the corresponding decision matrix of criterion values. 
Collecting these matrices yields the set 
$\{A^{(t')} : \text{phase}(t')=\phi\}$. For each criterion $j=1,\ldots,m$, we construct the empirical sample
 $\mathcal{X}_j = \{\, x_{ij}^{(t')} : t' \in \mathcal{K}_\phi,\;
i=1,\ldots,N \,\}$,
where $\mathcal{K}_\phi$ denotes the set of historical evaluation dates with phase $\phi$. This sample represents the historical behavior of the criterion $j$ under the same market conditions as those that prevail at the current evaluation date.

The investor’s preference structure is then incorporated through the two quantile-based 
thresholds
$
lp_{1,j} = Q_j(p_1)$ and $lp_{2,j} = Q_j(p_2)$, where $Q_j(\cdot)$ denotes the empirical quantile function associated with $\mathcal{X}_j$, and $(p_1,p_2)$ are the investor-specified percentiles. For cost criteria, the two thresholds are swapped to preserve the correct preference ordering. This process yields two dynamic limiting profiles $LP_1$ and $LP_2$ that adapt to the prevailing market conditions while reflecting the investor's preference structure.

Building on these definitions, the TODIMSort procedure unfolds in the following steps.
\begin{itemize}
\setlength{\itemsep}{0pt}
    \item \textit{Joint normalization.} 
    Let $\mathcal{A}$ denote the set of assets and 
    $\mathcal{P} = \{LP_1,LP_2\}$ the set of limiting profiles. 
    Form the combined set $\mathcal{U}=\mathcal{A}\cup\mathcal{P}$ and normalize 
    all criteria jointly to the interval $[0,1]$:
    \begin{equation}
        z''_{\alpha j} =
        \begin{cases}
            \dfrac{x_{\alpha j} - \min_{\gamma\in\mathcal{U}} x_{\gamma j}}
                  {\max_{\gamma\in\mathcal{U}} x_{\gamma j} - 
                   \min_{\gamma\in\mathcal{U}} x_{\gamma j}},
            & \text{if $j$ is a benefit criterion},\\[10pt]
            \dfrac{\max_{\gamma\in\mathcal{U}} x_{\gamma j} - x_{\alpha j}}
                  {\max_{\gamma\in\mathcal{U}} x_{\gamma j} -
                   \min_{\gamma\in\mathcal{U}} x_{\gamma j}},
            & \text{if $j$ is a cost criterion}.
        \end{cases}
    \end{equation}

    \item \textit{Local dominance.}
    For any pair $(\alpha,\beta)\in\mathcal{U}\times\mathcal{U}$ and any 
    criterion $j$, compute the local dominance contribution:
    \begin{equation}
        L_j(\alpha,\beta)=
        \begin{cases}
            \pi_j\,(z''_{\alpha j}-z''_{\beta j})^{\eta_1},
            & \text{if } z''_{\alpha j}\ge z''_{\beta j},\\[6pt]
            -\,\xi\,\pi_j\,(z''_{\beta j}-z''_{\alpha j})^{\eta_2},
            & \text{if } z''_{\alpha j} < z''_{\beta j}
        \end{cases}
    \end{equation}
    where $\pi_j$ are the MEREC weights and $(\eta_1,\eta_2,\xi)$ are the TODIM 
    sensitivity and loss-aversion parameters.

    \item \textit{Global dominance and scaling.}
    For each object $\alpha \in \mathcal{U}$, aggregate the local dominances 
    across all opponents and all criteria:
    \begin{equation}
        \Delta(\alpha)=
        \sum_{\beta\in \mathcal{U}}
        \sum_{j=1}^{m} L_j(\alpha,\beta).
    \end{equation}
    Then rescale dominance scores jointly over assets and profiles:
    \begin{equation}
        s(\alpha)=
        \frac{\Delta(\alpha)-\min_{\gamma\in\mathcal{U}}\Delta(\gamma)}
             {\max_{\gamma\in\mathcal{U}}\Delta(\gamma) -
              \min_{\gamma\in\mathcal{U}}\Delta(\gamma)}.
    \end{equation}

    \item \textit{Class assignment.}
    Let $l_1=s(LP_1)$ and $l_2=s(LP_2)$, with $0\le l_1\le l_2\le 1$. 
    Each asset $i\in\mathcal{A}$ is assigned to a class according to:
    \[
        s(i) < l_1 \ \Rightarrow\  i\in \text{short},\qquad
        l_1 \le s(i) < l_2 \ \Rightarrow\ i\in \text{neutral},\qquad
        s(i) \ge l_2 \ \Rightarrow\ i\in \text{long}.
    \]
\end{itemize}

This procedure yields an endogenous partition 
$(I_{\text{short}}, I_{\text{neutral}}, I_{\text{long}})$ whose cardinalities 
are entirely determined by the empirical dominance structure induced by the data. 
The resulting segmentation provides a preference-consistent and 
market-coherent investable universe, which serves as the initialization for the 
long--short portfolio optimization problem~\eqref{eq:omega_opt_long_short}.

\begin{remark}
The normalization adopted for MEREC follows the original protocol of the method, which requires strictly positive values due to the logarithmic aggregation of performance scores. By contrast, the TODIMSort procedure employs min–max normalization, which is more suitable for comparing alternatives and limiting profiles in a psychological ordering framework, and is fully consistent with the $[0, 1]$ scale required by the method. The use of different normalization schemes is, therefore, justified by the distinct nature and objectives of the two approaches.
\end{remark}

\subsection{Definition of Screening Criteria} \label{section:screeningCriteria}

The MEREC–TODIMSort architecture requires a structured set of quantitative criteria to build the decision matrices used in the weighting, dominance, and classification stages. These criteria integrate both structural features of assets and information that depends on the prevailing market environment, and are organized into two groups: phase‑independent indicators and phase‑dependent indicators.

\subsubsection{Market–Phase Identification}
\label{subsection:marketPhase}

A central component of this construction is the identification of the market phase, which determines when regime‑sensitive criteria are activated and governs the definition of the dynamic limiting profiles used by TODIMSort. To detect regime changes, we rely on a combination of standard technical indicators widely used in quantitative asset management. Their analytical definitions are reported in the Supplementary Material. Let $\{p_{b,t}\}_{t=1}^T$ denote the sequence of closing prices of the benchmark. The market phase at time $t$ is determined by combining three elementary signals: trend direction, short–term momentum, and trend strength.

\paragraph{(i) Trend direction} 
Define a short–term and a long–term moving average
\[
\mathrm{SMA}^{(S)}_t = \mathrm{SMA}(\{p_{b,\hat t}\}_{\hat t=1}^t,20), 
\qquad 
\mathrm{SMA}^{(L)}_t = \mathrm{SMA}(\{p_{b,\hat t}\}_{\hat t=1}^t,100)
\]
and let
$
\Delta_t = \mathrm{SMA}^{(S)}_t - \mathrm{SMA}^{(L)}_t
$
be their difference. To avoid spurious signals, we introduce a dynamic tolerance threshold proportional to the long–term level,
$\theta_t = 0.005\, \mathrm{SMA}^{(L)}_t$. A positive (resp.\ negative) deviation significantly above (resp.\ below) this threshold signals the presence of an upward (resp.\ downward) trend.

\paragraph{(ii) Short–term momentum}
Recent price momentum is captured by the $20$–day return
$r^{(20)}_{b,t} = (\frac{p_{b,t}}{p_{b,t-20}} - 1)$,
which reflects the instantaneous direction of the most recent movements.

\paragraph{(iii) Trend strength}
The Average Directional Index (ADX) is computed over a standard $14$–day horizon and denoted by $\mathrm{ADX}_t$. A threshold of $20$ is commonly adopted in practice to distinguish trending markets from low–directionality environments. Then, the condition $\mathrm{ADX}_t > 20$ indicates the presence of a trend.

\noindent Combining the above conditions, the market phase is classified as
\[
\mathrm{Phase}(t)=
\begin{cases}
\text{bull}, & \text{if } \Delta_t > \theta_t,\; r^{(20)}_{b,t} > 0,\; \mathrm{ADX}_t > 20, \\
\text{bear}, & \text{if } \Delta_t < -\theta_t,\; r^{(20)}_{b,t} < 0,\; \mathrm{ADX}_t > 20, \\
\text{sideways}, & \text{otherwise.}
\end{cases}
\]
The moving–average spread $\Delta_t$ captures the medium–term direction of the market, smoothing out high–frequency noise. The short–term return $r^{(20)}_{b,t}$ incorporates the most recent price dynamics, ensuring that the detected trend is not an artifact of past information. The ADX term provides a measure of directional strength: markets may exhibit positive or negative drift without being in a genuine trending state, and $\mathrm{ADX}_t$ filters out these situations. The combined rule thus balances robustness and responsiveness.

Although not directly used in the classification rule, we also compute a volatility indicator based on the relative magnitude of short‑ and long‑term benchmark return variability. Specifically, $\sigma^{(20)}_t = \operatorname{std}\!\left(r_{t-19},\dots,r_t\right)$ and $\sigma^{(\mathrm{all})}_t = \operatorname{std}\!\left(r_1,\dots,r_t\right)$ denote the 20‑day realized volatility of the benchmark and its long‑horizon (3‑year) realized volatility, respectively. Then, we calculate the volatility ratio as $\varrho_t = \nicefrac{\sigma^{(20)}_t}{\sigma^{(\mathrm{all})}_t}$. Finally, we classify the volatility regime as high if $\varrho_t > 1.2$, as low if $\varrho_t < 0.8$, and as normal otherwise. The volatility classification complements the market‑phase signal and plays a direct role in defining the regime‑concordant technical score introduced in the sequel.

\subsubsection{Phase–Independent Criteria}
Let $\{r_{i,t}\}$ denote the return series of asset $i$ and $\{r_{b,t}\}$ the benchmark returns. Phase–independent criteria point out structural properties of assets that are informative across market regimes.

\noindent\emph{Momentum.}
Medium–term momentum measures recent compounding performance and is a robust predictor of cross–sectional returns \citep{DanielMoskowitz2016}. Over a lookback window of length $H_{\text{mom}}$, we define
\[
\mathrm{MOM}_{i,t}
= \prod_{h=0}^{H_{\text{mom}}-1}\!\bigl(1+r_{i,t-h}\bigr) - 1.
\]
Assets with higher momentum values are considered more attractive.

\noindent\emph{Upside–to–downside beta ratio.}
It measures the asymmetric responsiveness of asset $i$ to positive and negative benchmark movements \citep{FisherDalessandro2021}. Let $\tau$ be a state threshold. Over a lookback window of length $H_{\text{udr}}$, we define the state partitions $\mathcal{T}^{+} = \{u:\, r_{b,u} > \tau\}$ and $\mathcal{T}^{-} = \{u:\, r_{b,u} < \tau\}$. For each asset $i$, compute the conditional betas
\[
\beta^{+}_{i}
=
\frac{\mathrm{Cov}\!\bigl(r_{i,u},r_{b,u}\mid u\in\mathcal{T}^{+}\bigr)}
     {\mathrm{Var}(r_{b,u}\mid u\in\mathcal{T}^{+})},
\qquad
\beta^{-}_{i}
=
\frac{\mathrm{Cov}\!\bigl(r_{i,u},r_{b,u}\mid u\in\mathcal{T}^{-}\bigr)}
     {\mathrm{Var}(r_{b,u}\mid u\in\mathcal{T}^{-})}
\]
and the asymmetry ratio $\mathrm{UDR}_{i,t}^{\tau} = \frac{\beta^{+}_{i}}{\beta^{-}_{i}}.$ 
Higher values of UDR indicate that the asset tends to load more strongly on favorable benchmark states than on unfavorable ones, reflecting a desirable form of asymmetry in risk–return behavior.

\noindent\emph{Drawdown–recovery metrics.}
Let $\{p_{i,u}\}$ denote the price series of asset $i$ over a lookback window of length $H_{\text{dd}}$, and define the running peak and drawdown processes as
\[
P^{\max}_{i,u} = \max_{k \le u} p_{i,k},
\qquad
\mathrm{DD}_{i,u} = 1 - \frac{p_{i,u}}{P^{\max}_{i,u}}.
\]
In this study, we employ two recovery–based downside–risk indicators \citep{Chekhlov2005}: recovery time and recovery factor. A drawdown is initiated at index $u$ whenever $p_{i,u} < P^{\max}_{i,u}$.  
For such a drawdown, its recovery time is defined as $m(u) = \min\{ m > u : p_{i,m} \ge P^{\max}_{i,u} \}$, i.e.,
 the first time the price returns to its previous peak.  
Let $\mathcal{U}$ denote the set of drawdown onset points that recover within the window. Average recovery time is then
\[
\mathrm{RT}_{i,t}
= 
\frac{1}{|\mathcal{U}|}
\sum_{u \in \mathcal{U}}
\bigl(m(u)-u\bigr).
\]
Lower values of $\mathrm{RT}_{i,t}$ indicate faster rebounds following adverse movements and are therefore preferred.

Let $\mathrm{MaxDD}_{i,t} = \max_{1 \le u \le H_{\text{dd}}}
\mathrm{DD}_{i,u}$ denote the maximum drawdown over the same window, and define the cumulative return as $\mathrm{CR}_{i,t}
=
\prod_{u=1}^{H_{\text{dd}}}
\left(1 + r_{i,u}\right)
- 1$. The recovery factor is then given by
\[
\mathrm{RF}_{i,t}
=
\frac{\mathrm{CR}_{i,t}}{\mathrm{MaxDD}_{i,t}}.
\]
Higher values of $\mathrm{RF}_{i,t}$ correspond to more efficient compensation of losses and are therefore preferred.

\noindent\emph{ESG–based indicators.}
Environmental, Social, and Governance assessments provide complementary information on firms’ non‑financial performance and long‑term sustainability. Let $\{\mathrm{ESG}_{i,u}\}$ denote the monthly ESG score series of asset $i$. Given a lookback parameter $H_{\mathrm{esg}}$, we consider two classes of indicators \citep{Nagy2016}: ESG momentum and ESG volatility. The ESG momentum of asset $i$ over a horizon of length $H_{\mathrm{esg}}$ is defined as
$\mathrm{ESG\text{-}MOM}_{i,t}^{H_{\mathrm{esg}}}
    = 
\mathrm{ESG}_{i,t}
-
\mathrm{ESG}_{i,t-H_{\mathrm{esg}}}$. This indicator captures the direction and magnitude of recent changes in a firm’s sustainability profile. Higher values of $\mathrm{ESG\text{-}MOM}_{i,t}$ reflect improvements in ESG performance and are therefore preferred. ESG volatility measures the stability of a firm’s sustainability performance over the same horizon, and is defined as
\[
\mathrm{ESG\text{-}VOL}_{i,t}^{H_{\mathrm{esg}}}
=
\sqrt{
\frac{1}{H_{\mathrm{esg}}}
\sum_{h=0}^{H_{\mathrm{esg}}-1}
\bigl(
\mathrm{ESG}_{i,t-h}
-
\overline{\mathrm{ESG}}_{i,t}
\bigr)^2}
\]
where $\overline{\mathrm{ESG}}_{i,t}
=
\frac{1}{H_{\mathrm{esg}}}
\sum_{h=0}^{H_{\mathrm{esg}}-1}
\mathrm{ESG}_{i,t-h}$.
Lower values of $\mathrm{ESG\text{-}VOL}_{i,t}$ indicate a more stable ESG behavior and are therefore preferred.

\subsubsection{Phase–Dependent Criteria} \label{subsection:depententCriteria}
These indicators adapt the screening to the prevailing regime $\phi\in\{\text{bull},\text{bear},\text{sideways}\}$ detected by the market–phase identifier. They modulate diversification and tracking behaviour in a manner consistent with current conditions.

\noindent\emph{Normalized Mutual Information with the benchmark.}
Let $\{r_{i,u}\}$ and $\{r_{b,u}\}$ denote, respectively, the daily return series of asset $i$ and of the benchmark over a lookback window of length $H_{\text{nmi}}$. To quantify the degree of dependence between each asset and the market, we compute the normalized mutual information (NMI) between the return series of the asset and the benchmark:
\[
\mathrm{NMI}_{i,t}
=
\frac{2\,MI(r_{i};r_{b})}
     {H^{\text{Sh}}(r_{i})+H^{\text{Sh}}(r_{b})}
\in[0,1]
\]
where $MI(\cdot,\cdot)$ denotes mutual information and $H^{\text{Sh}}(\cdot)$ Shannon entropy (their definitions are provided in the Supplementary Material). This formulation symmetrically normalizes the dependence measure, ensuring comparability across assets and preventing scale distortions due to heterogeneous return distributions. For each asset, $\mathrm{NMI}_{i,t}$ is estimated non-parametrically using the procedure described in \cite{kraskov2004}. To incorporate regime–specific preferences into the NMI measure, we transform the score 
according to the prevailing phase:
\[
\widetilde{\mathrm{NMI}}_{i,t}=
\begin{cases}
\mathrm{NMI}_{i,t},
& \phi_t\in\{\text{bull},\text{bear}\}, \\[4pt]
1-\lvert \mathrm{NMI}_{i,t}-0.5\rvert,
& \phi_t=\text{sideways}.
\end{cases}
\]
In bull markets, strong co-movement with the benchmark is desirable, so NMI is used as a benefit criterion. In bear markets, high dependence makes the asset more exposed to market downturns; 
therefore, NMI is marked as a cost criterion. In sideways regimes, neither very high nor very low co-movement is optimal, as range–bound environments reward moderately correlated behavior. Transformation $1-\lvert \mathrm{NMI}_{i,t}-0.5\rvert$ achieves this by assigning the highest score to values near $0.5$ and penalizing extreme co-movement.  The transformed quantity is then treated as a benefit criterion.

\noindent\emph{Regime‑concordant technical signals.}
We construct for each asset a composite regime–concordant technical score $S^{\mathrm{tech}}_{i,t}$, designed to summarize short– and medium–term price information in a manner that reflects the prevailing market phase $\phi_t\in\{\text{bull},\text{bear},\text{sideways}\}$ 
and its associated volatility condition $\upsilon_t\in\{\text{low},\text{normal},\text{high}\}$. The score is obtained by aggregating the outputs of phase–specific signal functions applied to asset price series $\{p_{i,u}\}$ over a lookback window of  $H_{\text{tech}}$ days, each relying on a set of standard technical indicators whose operational definitions are reported in the Supplementary Material. For each asset $i$, the signaling rule returns a raw indicator score $\mathrm{sc}_{i,t}$, which we subsequently transform into a continuous measure $S^{\mathrm{tech}}_{i,t}\in[-1,1]$ suitable for integration within the MEREC–TODIMSort machinery. In trending regimes ($\phi_t\in\{\text{bull},\text{bear}\}$), each asset is 
evaluated through three elementary components: (i) trend strength via ADX, (ii) directional momentum via MACD, and (iii) a momentum–confirmation filter based on RSI. Let $\mathrm{ADX}_{i,t}$, $\mathrm{MACD}_{i,t}$, and $\mathrm{RSI}_{i,t}$ denote the corresponding indicator values. The score $\mathrm{sc}_{i,t}\in\{0,1,2,3\}$ counts how many of the following conditions are satisfied:
\begin{enumerate}
\setlength{\itemsep}{0pt}
\item $\mathrm{ADX}_{i,t} > 20$ (trend strength);
\item $\mathrm{MACD}_{i,t} > 0$ in bull markets, 
      $\mathrm{MACD}_{i,t} < 0$ in bear markets (directional momentum);
\item $\mathrm{RSI}_{i,t} \in [30, 70]$ (momentum confirmation).
\end{enumerate}

To obtain a unified directional scale, the raw score is mapped to $[-1,1]$ as
\[
S^{\mathrm{tech}}_{i,t}=
\begin{cases}
\displaystyle\frac{\mathrm{sc}_{i,t}-1.5}{1.5}, 
& \phi_t=\text{bull}, \\[6pt]
\displaystyle-\frac{\mathrm{sc}_{i,t}-1.5}{1.5}, 
& \phi_t=\text{bear}
\end{cases}
\]
so that higher values always favor long positions and lower values favor short positions, independently of the trend direction.

When the market is sideways and volatility is high ($\phi_t=\text{sideways},\ \upsilon_t=\text{high}$), the scoring rule targets range–bound dynamics with large excursions, using Bollinger–band signals, the Commodity Channel Index (CCI), and RSI. Let $\mathrm{BBsig}_{i,t}\in\{-1,0,1\}$, $\mathrm{CCI}_{i,t}$, and $\mathrm{RSI}_{i,t}$ represent the signals obtained from these indicators. We compute two partial scores, $\mathrm{buySc}_{i,t}$ and $\mathrm{sellSc}_{i,t}$, by counting the number of indicators simultaneously supporting oversold or overbought conditions, respectively, using the thresholds $\mathrm{CCI}\in[-100,100]$ and $\mathrm{RSI}\in[30,70]$. The continuous technical score is then defined as
\[
S^{\mathrm{tech}}_{i,t}=
\frac{\mathrm{buySc}_{i,t}-\mathrm{sellSc}_{i,t}}{3},
\qquad 
S^{\mathrm{tech}}_{i,t}\in[-1,1].
\]

When the market is sideways with low or normal volatility 
($\phi_t=\text{sideways},\ \upsilon_t\in\{\text{low},\text{normal}\}$), we adopt 
a finer range–trading logic based on the RSI, the Stochastic Oscillator ($K$–$D$ crossing), and the CCI. Let $\mathrm{RSI}_{i,t}$, 
$K_{i,t}$, $D_{i,t}$, $\mathrm{CCI}_{i,t}$ be the corresponding signal values, buy–oriented evidence is recorded when $\mathrm{RSI}_{i,t}<30$, $K_{i,t}>D_{i,t}$, and $\mathrm{CCI}_{i,t}<-100$; sell–oriented evidence when $\mathrm{RSI}_{i,t}>70$, $K_{i,t}<D_{i,t}$, and $\mathrm{CCI}_{i,t}>100$. Letting $\mathrm{buySc}_{i,t}$ and $\mathrm{sellSc}_{i,t}$ denote the associated counts, the continuous regime–concordant score is
\[
S^{\mathrm{tech}}_{i,t}=
\frac{\mathrm{buySc}_{i,t}-\mathrm{sellSc}_{i,t}}{3}
\in[-1,1].
\]

\section{Optimization Framework} 
\label{section:OptimizationAlgorithm}

We now introduce the adaptive multi-operator particle swarm optimization algorithm with projection-based repair mechanism (shortly, AMPSO), designed to solve the proposed portfolio optimization problems. The algorithm integrates three components: 
\begin{itemize}
\setlength{\itemsep}{0pt}
    \item[(i)] a particle swarm optimization with time-varying acceleration coefficients (PSO-TVAC) serving as the core search engine;
    \item[(ii)] an adaptive operator selection strategy (AOS) that dynamically combines PSO dynamics with a set of variation operators based on performance feedback;
    \item[(iii)] a constraint-repair technique that ensures feasibility with portfolio constraints.
\end{itemize}
Although our portfolio optimization problems are formulated as maximization of the Omega ratio, in practice we solve equivalent minimization problems by minimizing the function $f(w) = -\Omega(R_w; R_b)$ while keeping the feasible regions unchanged. This transformation is standard in the metaheuristic optimization literature, where solvers are typically designed to minimize objective functions.

\subsection{PSO Algorithm with Time-Varying Acceleration Coefficients}\label{subsection:PSOTVAC}

Particle swarm optimization is a population-based stochastic search method inspired by social learning and collective behavior. In its canonical form, PSO maintains a swarm of particles, each representing a candidate solution in an $n$-dimensional search space. At iteration $g = 0,1,\dots,g_{\max}$, particle $j$ has position $y_j(g) \in \mathbb{R}^n$ and velocity $v_j(g) \in \mathbb{R}^n$. The update mechanism consists of two coupled equations:
\begin{align}
v_j(g+1) &= \omega(g)\,v_j(g) 
+ c_1(g)\,r_1(g)\odot\big(y_j^\star(g) - y_j(g)\big)
+ c_2(g)\,r_2(g)\odot\big(\hat y(g) - y_j(g)\big) \label{eq:velocity_update}\\
y_j(g+1) &= y_j(g) + v_j(g+1) \label{eq:position_update}
\end{align}
where $\odot$ denotes the element-wise product, $r_1(g), r_2(g) \sim U(0,1)^n$ are random vectors, with each component independently drawn from the uniform distribution on $[0,1]$, and $y_j^\star(g)$ and $\hat y(g)$ represent the best positions found by particle $j$ and by the entire swarm, respectively.

The inertia weight $\omega(g)$ regulates the trade-off between global exploration and local exploitation by scaling the contribution of the previous velocity. Large values of $\omega(g)$ favor exploration, while small values promote convergence. Similarly, the acceleration coefficients $c_1(g)$ and $c_2(g)$ control the cognitive and social components, weighting the attraction towards the particle's own best position and the global best position, respectively.

To achieve a dynamic balance between exploration and exploitation, the PSO-TVAC algorithm employs linearly adaptive schedules (see \citet{Ratnaweera2004}, and \citet{Deng2012} for an application in the field of portfolio selection):
\begin{align}
\omega(g) &= \omega_{\max} - \frac{\omega_{\max} - \omega_{\min}}{g_{\max}}\,g, \label{eq:TVAC1} \\
c_1(g) &= c_{1,\max} - \frac{c_{1,\max} - c_{1,\min}}{g_{\max}}\,g, \label{eq:TVAC2}\\
c_2(g) &= c_{2,\min} + \frac{c_{2,\max} - c_{2,\min}}{g_{\max}}\,g \label{eq:TVAC3}
\end{align}
where $\omega_{\max} > \omega_{\min}$, $c_{1,\max} > c_{1,\min}$, and $c_{2,\max} > c_{2,\min}$. 

In our version of the algorithm, to prevent divergence and maintain stability, velocities are clamped component-wise 
$v_j(g+1) \leftarrow \min\big\{\max\big(v_j(g+1), v_{\min}\big), v_{\max}\big\}$, where $v_{\min}$ and $v_{\max}$ denote the component-wise lower and upper velocity bounds \citep{ShiEberhart1998}. 

The PSO-TVAC procedure can thus be summarized as follows:
\begin{enumerate}
\setlength{\itemsep}{0pt}
\item Initialize particle positions uniformly within the search bounds and set initial velocities to zero.
\item Evaluate the objective function for all particles; initialize $y_j^\star$ and $\hat y$.
\item For $g = 1,\dots,g_{\max}$:
    \begin{enumerate}
    \setlength{\itemsep}{0pt}
    \item Update $\omega(g)$, $c_1(g)$, and $c_2(g)$ according to the TVAC schedules \eqref{eq:TVAC1}--\eqref{eq:TVAC3}.
    \item Compute new velocities using Eq.~\eqref{eq:velocity_update} and apply velocity clamping.
    \item Update positions using Eq.~\eqref{eq:position_update}.
    \item Evaluate the objective function and update $y_j^\star$ and $\hat y$.
    \end{enumerate}
\item Return $\hat y$ as the best solution found.
\end{enumerate}

The adaptation in time of $\omega$, $c_1$, and $c_2$ promotes global search during early iterations and accelerates convergence in later stages, mitigating the risk of getting trapped in local optima and improving the quality of solutions in high-dimensional, non‑convex landscapes, such as long–short portfolio optimization.

\subsection{Adaptive Operator Selection Strategy}

The proposed metaheuristic employs an adaptive operator selection (AOS) mechanism to regulate the balance between exploration and exploitation. AOS operates on a heterogeneous set of variation operators, including real-coded crossovers, multi-parent recombination schemes, discrete operators, and mutation mechanisms. Detailed definitions and parameter settings for all operators are reported in the  Supplementary Material. At each generation, AOS evaluates the relative performance 
of the available operators and probabilistically selects the one expected to offer  the best trade-off between solution quality and population diversity. The controller  follows a four-stage reinforcement-learning scheme consisting of criteria aggregation, reward computation, credit assignment, and operator selection.

\paragraph{Criteria aggregation}
At generation $g$, a subset of $n_{\mathrm{par}}$ particles is sampled and each 
operator in $\mathcal{O}=\{o_1,\ldots,o_{n_{\mathrm{op}}}\}$ is applied to this 
subset, generating trial offspring. 
For operator $o_k$, we record 
 $Q_g^{(k)}$ (i.e., the mean objective value of the offspring) and $D_g^{(k)}$ (i.e., the mean Euclidean dispersion of the offspring), 
where dispersion is measured as the average distance from the population centroid.\footnote{Recall that throughout this section all performance statistics are computed with respect to the minimization objective $f(w) = -\Omega(R_w; R_b)$.} To reduce short-term noise, improvements are aggregated over a sliding window of 
length $T_\text{w}$:
\[
\Delta_Q^{(k)}(g)=\frac{1}{T_\text{w}-1}\sum_{t=g-T_\text{w}+1}^{g-1}
\bigl(Q_{t+1}^{(k)} - Q_t^{(k)}\bigr),
\qquad
\Delta_D^{(k)}(g)=\frac{1}{T_\text{w}-1}\sum_{t=g-T_\text{w}+1}^{g-1}
\bigl(D_{t+1}^{(k)} - D_t^{(k)}\bigr).
\]

\paragraph{Reward computation}
Rewards are computed using the compass method, which evaluates the alignment of the vector $\left(\Delta_Q^{(k)},\Delta_D^{(k)}\right)$ with a target search direction parameterized by an angle $\varphi_g\in[0,\pi/2]$ \citep{filograsso2023}. The controller alternates exploration and exploitation through the schedule

\begin{equation}\label{phiAOS}
\varphi_g=
\begin{cases}
\frac{\pi}{2}, & g\le 0.2\,g_{\max},\\
0,             & 0.2\,g_{\max} < g \le 0.4\,g_{\max},\\
\frac{\pi}{2}, & 0.4\,g_{\max} < g \le 0.6\,g_{\max},\\
0,             & 0.6\,g_{\max} < g \le 0.8\,g_{\max},\\
\frac{\pi}{2}, & g > 0.8\,g_{\max}.
\end{cases}
\end{equation}

Let $m=\tan(\varphi_g)$ be the slope of the target direction in the $(\Delta_Q,\Delta_D)$ plane. For operator $o_k$ we compute $d = \sqrt{\bigl(\Delta_Q^{(k)}\bigr)^2 + \bigl(\Delta_D^{(k)}\bigr)^2}$, and $d_\perp = \frac{\bigl|\Delta_D^{(k)} - m\,\Delta_Q^{(k)}\bigr|}{\sqrt{1+m^2}}$. Then, the quantity ${reward}^{(k)}$ is computed as $\sqrt{d^2 - d_\perp^2}$, which corresponds to the projection of $\left(\Delta_Q^{(k)},\Delta_D^{(k)}\right)$ onto the 
target direction.

\paragraph{Credit assignment}
Credits are obtained by averaging rewards over the sliding window
\[
c_k=\frac{1}{T_\text{w}}\sum_{t=g-T_\text{w}+1}^{g}{reward}^{(k)}(t).
\]

\paragraph{Operator selection}
The next operator is selected by probability matching:
\[
p_k = p_{\min} + 
(1-n_{\mathrm{op}}\,p_{\min}) 
\frac{c_k}{\sum_{j=1}^{n_{\mathrm{op}}} c_j}\,,
\]
where $p_{\min}>0$ ensures non-zero probability for all operators. The operator corresponding to the highest $p_k$ is applied to each particle with probability $p_{\mathrm{app}}$. This mechanism enables the algorithm to favor operators that consistently improve quality when $\varphi_g=\pi/2$ and those that increase diversity when $\varphi_g=0$, thus maintaining a robust balance between exploration and exploitation throughout the search.

\subsection{Constraint Handling via Projection Methods}
\label{subsection:CHT}

The portfolio optimization problems introduced in Section~\ref{section:PortfolioModel} involve equality and 
bound constraints that define a closed and convex feasible region. To guarantee 
that every particle generated by variation operators remains feasible, we adopt a 
projection-based repair strategy that maps any infeasible vector to its 
orthogonal projection onto the constraint set. 
This approach preserves feasibility without altering the objective function and 
is well suited to linear budget constraints coupled with box bounds.

\subsubsection{Projection for the long-short model}
\label{subsection:CHT}

After classifying the assets into the two disjoint index sets 
$I_{\text{long}}$ and $I_{\text{short}}$, the feasible region of the long--short 
problem~\eqref{eq:omega_opt_long_short} is
\[
S = 
\biggl\{
w\in\mathbb{R}^n :
\sum_{i\in I_{\text{long}}} w_i = 1+s,\;
0\le w_i\le w_i^{u}\;\forall i\in I_{\text{long}},
\;
\sum_{i\in I_{\text{short}}} w_i = -s,\;
w_i^{\ell}\le w_i\le 0\;\forall i\in I_{\text{short}}
\biggr\}.
\]
Since $I_{\text{long}}\cap I_{\text{short}}=\emptyset$ and the Euclidean norm is 
separable, the projection of a vector $e_w\in\mathbb{R}^n$ onto $S$ decomposes 
into two independent subproblems:
\[
\mathrm{PS}(e_w)
=
\bigl(
\mathrm{PS}_L(e_{w,(L)}),\;
\mathrm{PS}_S(e_{w,(S)})
\bigr)
\]
where $e_{w,(L)}=(e_{w,i})_{i\in I_{\text{long}}}$ and 
$e_{w,(S)}=(e_{w,i})_{i\in I_{\text{short}}}$.

\paragraph{Long block}
The projection for the long block solves
\[
\min_{u\in\mathbb{R}^{n_{\text{long}}}}
\frac{1}{2}\|u - e_{w,(L)}\|_2^2
\quad\text{s.t.}\quad
\sum_{i\in I_{\text{long}}} u_i = 1+s,\;
0\le u_i\le w_i^{u}.
\]
By the Karush--Kuhn--Tucker conditions, the optimal solution has the form
\begin{equation} \label{Eqn:Proj1}
u_i = 
\bigl[e_{w,i} - \lambda \bigr]^{\,w_i^{u}}_{\,0},
\qquad i\in I_{\text{long}}    
\end{equation}
where $[\theta]^{b}_{a}=\min\{\max\{\theta,a\},b\}$ denotes clipping, and $\lambda$ is the unique scalar satisfying $
\sum_{i\in I_{\text{long}}}
\bigl[e_{w,i} - \lambda \bigr]^{\,w_i^{u}}_{\,0}
=1+s.$

\paragraph{Short block}
Similarly, the short block projection solves
\[
\min_{v\in\mathbb{R}^{n_{\text{short}}}}
\frac{1}{2}\|v - e_{w,(S)}\|_2^2
\quad\text{s.t.}\quad
\sum_{i\in I_{\text{short}}} v_i = -s,\;
w_i^{\ell}\le v_i\le 0
\]
whose solution is
\begin{equation} \label{Eqn:Proj2}
v_i =
\bigl[e_{w,i} - \lambda' \bigr]^{\,0}_{\,w_i^{\ell}},
\qquad i\in I_{\text{short}}
\end{equation}
with $\lambda'$ uniquely determined by
$  
\sum_{i\in I_{\text{short}}}
\bigl[e_{w,i} - \lambda' \bigr]^{\,0}_{\,w_i^{\ell}}
= -s.
$   

Existence and uniqueness of $\lambda$ and $\lambda'$ follow from the fact that the left-hand sides of the above equations define continuous, strictly decreasing functions whose ranges cover the imposed budgets. In practice, the values of $\lambda$ and $\lambda'$ can be efficiently computed using the $O(n)$ algorithm for projecting a vector onto the intersection of a hyperplane and a box, as described in \citet{maculan2003}. This guarantees a numerically stable and computationally scalable implementation of operator PS.

\subsubsection{Long-only model}

Under short-selling suspension, the feasible region reduces to
\[
S' = 
\biggl\{
w\in\mathbb{R}^{n_{\text{long-only}}} :
\sum_{i\in I_{\text{long-only}}} w_i = 1,\;
0\le w_i\le w_i^{u}
\biggr\}.
\]
The projection onto $S'$ is the specialization of the long-block projector \eqref{Eqn:Proj1}:
\[
w_i = \bigl[e_{w,i} - \lambda \bigr]^{\,w_i^{u}}_{\,0},
\qquad i\in I_{\text{long-only}}
\]
where $\lambda$ is chosen so that 
$\sum_{i\in I_{\text{long-only}}}
\bigl[e_{w,i} - \lambda \bigr]^{\,w_i^{u}}_{\,0} = 1$.

\subsection{AMPSO}

The AMPSO solver integrates the components introduced in Sections~\ref{subsection:PSOTVAC}-\ref{subsection:CHT} into a unified multi‑operator swarm framework. At each iteration, particles evolve according to the PSO–TVAC dynamics and are subsequently projected onto the feasible region to enforce budget and long–short constraints. The adaptive operator selection (AOS) mechanism then complements the PSO update by choosing, with probability matching, the variation operator that has shown the most favorable balance between quality improvement and population dispersion over a sliding performance window.

This integration allows AMPSO to alternate between exploitation phases, dominated by PSO–TVAC updates, and exploration phases, driven by the adaptive activation of 
mutation and crossover operators. The overall procedure yields a solver capable of adjusting its search behavior in response to the topology of the optimization 
landscape while maintaining feasibility at every iteration. A complete pseudocode of AMPSO, together with the full specification of the variation operators, is 
reported in the Supplementary Material.

\section{Experimental Analysis} \label{section:experiments}

This section presents the empirical assessment of the proposed long–short portfolio optimization framework. First, we introduce the data set used in the study and the parameters adopted for the screening criteria. Second, we examine the effectiveness of the MEREC–TODIMSort procedure in assigning assets to long, neutral, and short categories. Third, we evaluate the performance of the AMPSO algorithm and compare its behavior with standard evolutionary benchmarks. Finally, we assess the ex‑post profitability of the portfolios generated by the full pipeline.

\subsection{Data and Parameter Specification} \label{subsection:dataDescription}

The empirical analysis is conducted on a European equity universe composed of a subset of the constituents of the STOXX Europe~600 index. Daily closing prices, market capitalization, and ESG scores are retrieved from Refinitiv, and all series are aligned across the sample to ensure temporal consistency. The data set spans the period from 1~June~2013 to 30~June~2023. Assets with missing observations within any relevant estimation window are removed, resulting in an investment universe of $N = 421$ stocks.

The backtesting procedure follows a monthly rebalancing scheme. On each rebalancing date, defined as the last trading day of the month, we utilize a 3-year rolling in-sample window to obtain the historical information needed to estimate the empirical Omega ratio and to compute the quantities involved in the screening and classification stages. Then, we allocate the optimal portfolio to the subsequent 1-month out-of-sample period, in line with the single-period framework described in Section~\ref{section:PortfolioModel}. In addition to price-based information, the pre-selection stage requires the construction of regime-conditioned decision matrices used to define the dynamic limiting profiles of the TODIMSort procedure (Section~\ref{subsection:TODIMSort}). Since at least two historical matrices are needed for each market phase, we impose a minimum of 12 initial months before the first evaluation date. Combined with the 3-year rolling window, this implies that the empirical study is carried out on 72 admissible rebalancing dates, from 31~July~2017 to 30~June~2023. This trimming step ensures that all computations rely on statistically coherent price histories and regime-specific information sets that remain representative of the prevailing market conditions.

The data set also incorporates relevant regulatory constraints. During the COVID-19 turmoil, the Italian market authority CONSOB introduced a temporary ban on creating or increasing net short positions. The measure, enacted by Resolution No. 21303 on March 17, 2020 and endorsed by ESMA, remained in force from March 18 to June 18, 2020. Hence, we treated the rebalancing windows intersecting March, April, and May 2020 as long-only, ensuring full adherence to the regulatory environment.

The implementation of the MEREC–TODIMSort framework requires a consistent parameter specification for all screening criteria, that we report in Table \ref{tab:criteria_parameters}. Medium-term momentum is computed using monthly returns over a 1-year lookback horizon, with a 1-month lag to mitigate short-term reversal effects. The upside-to-downside beta ratio is estimated using two alternative benchmark state thresholds $\tau \in \bigl\{ q_{0.25}(r_b),\; q_{0.50}(r_b) \bigr\}$, where $q_{\delta}(r_b)$ denotes the empirical $\delta$-quantile of benchmark returns. The threshold $\tau = q_{0.25}(r_b)$ captures adverse market events, while $\tau = q_{0.50}(r_b)$ provides a balanced state partition consistent with typical market conditions. Drawdown-related indicators are evaluated using an daily observations on a 3-year in-sample window. ESG-based indicators are computed from end-of-month Refinitiv scores: ESG momentum uses a 12-month horizon, while ESG volatility is computed over 24- and 36-month windows, with a 1-month lag applied uniformly to prevent look-ahead bias. Regime-dependent indicators, including Normalized Mutual Information (NMI) and regime-concordant technical score $S_{\text{tech}}$, rely on the market-phase detection mechanism of Section \ref{subsection:marketPhase} and therefore inherit the same rolling estimation windows for return- and price-based inputs. This ensures that all regime-conditioned decision matrices remain synchronized and constructed exclusively from information available at each rebalancing date.

\begin{table}[t]
    \centering
    \footnotesize
    \caption{Summary of screening criteria and parameter settings used in the MEREC--TODIMSort framework.}
    \label{tab:criteria_parameters}
    \begin{tabularx}{\textwidth}{l l l l X}
        \hline
        \textbf{Criterion} & \textbf{Symbol} & \textbf{Lookback} & \textbf{Type} \\ 
        \hline
        Momentum 
            & $\mathrm{MOM}_i$
            & 1 year (monthly) and 1-month lag 
            & Benefit \\

        Upside--Downside Beta Ratio \\
        \hspace*{1em}(0.25 quantile of benchmark returns)
            & $\mathrm{UDR}^{0.25}_i$
            & 2 years (daily)
            & Benefit \\

        Upside--Downside Beta Ratio \\
        \hspace*{1em}(0.50 quantile of benchmark returns)
            & $\mathrm{UDR}^{0.50}_i$
            & 2 years (daily)
            & Benefit \\
            
        Recovery Time
            & $\mathrm{RT}_i$
            & 3 years (daily)
            & Cost \\

        Recovery Factor
            & $\mathrm{RF}_i$
            & 3 years (daily)
            & Benefit \\

        ESG Momentum
            & $\mathrm{ESG\text{-}MOM}_i$
            & 1 year (monthly) and 1-month lag 
            & Benefit \\

        ESG Volatility (24m)
            & $\mathrm{ESG\text{-}VOL}^{24}_i$
            & 2 years (monthly) and 1-month lag
            & Cost \\
            
        ESG Volatility (36m)
            & $\mathrm{ESG\text{-}VOL}^{36}_i$
            & 3 years (monthly) and 1-month lag
            & Cost \\

        Normalized Mutual Information
            & $\mathrm{NMI}_i$
            & 2 years (daily)
            & Phase-dependent \\

        Regime-Concordant Technical Score
            & $S_{\mathrm{tech},i}$
            & Indicator-specific
            & Benefit \\
        \hline
    \end{tabularx}
\end{table}

For benchmarking purposes, a synthetic market-value-weighted index constructed from the 421 assets in the investment universe is used as the market proxy throughout the empirical analysis.

\subsection{Analysis of the MEREC--TODIMSort Procedure}
\label{subsec:merec_todimsort_analysis}

This subsection examines the behavior of the screening and classification component, detailing how regime identification, MEREC weighting patterns, and TODIMSort limiting profiles jointly shape the construction of long, neutral, and short sets.

\subsubsection{Market-phase identifier}

\begin{figure}[t]
\centering
\includegraphics[width=0.70\textwidth]{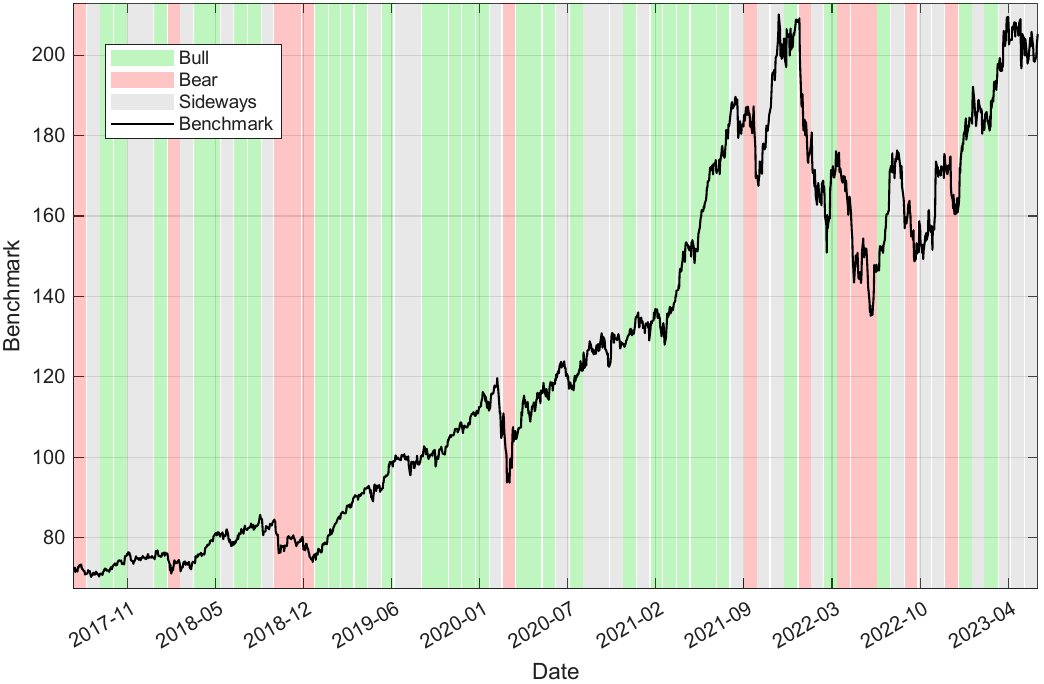}
\caption{Evolution of the benchmark index over the experimental period with soft‑colored bands indicating the market phase detected at each rebalancing date (bull, bear, sideways). Regime labels are based exclusively on information available up to the corresponding date.}
\label{fig:phaseIdentifier}
\end{figure}

A central component of the screening and classification architecture is the market-phase identifier, which governs both the activation of regime-dependent criteria (such as NMI and the regime-concordant technical score) and the construction of the limiting profiles used in TODIMSort. Its reliability is therefore essential to ensure that the criteria matrix and the class boundaries remain consistent with prevailing market conditions. Figure~\ref{fig:phaseIdentifier} shows the benchmark together with the phases detected at each rebalancing date. The alternation between bull, bear, and sideways states aligns closely with observable shifts in trend direction, short‑term momentum, and directional strength, indicating that the identifier captures regime changes in a timely and coherent manner. This alignment is obtained without anticipatory information: all labels are produced under a strict no–look–ahead policy and rely exclusively on data available at each evaluation date. The evidence in Figure \ref{fig:phaseIdentifier} indicates that a classical yet suitably filtered set of technical indicators can generate a parsimonious and stable regime map, providing a coherent foundation for regime‑conditioned screening and classification decisions.

\subsubsection{MEREC: regime-dependent relevance of screening criteria}

\begin{figure}[t]
\centering
\includegraphics[width=0.70\textwidth]{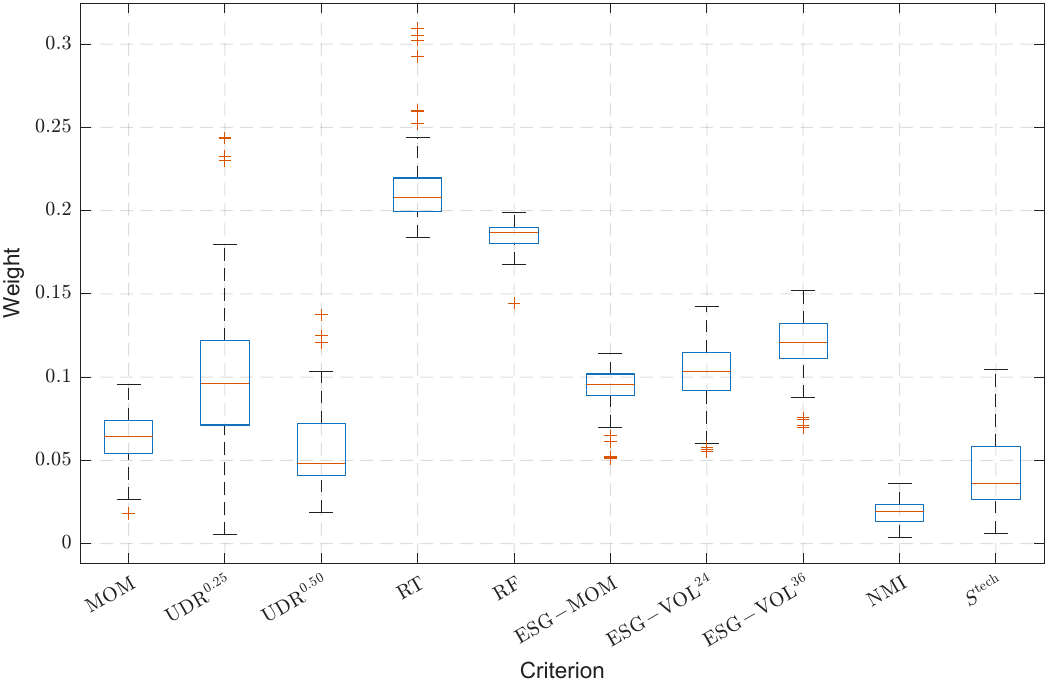}
\caption{Box plot of the criterion weights assigned by the MEREC method across all rebalancing periods.}
\label{fig:boxplotMEREC}
\end{figure}

Next, we examine how MEREC allocates importance across criteria and how this allocation varies with market conditions. 
From the joint inspection of the box plots in Figure~\ref{fig:boxplotMEREC} and the phase-specific averages reported in Table~\ref{tab:MEREC_summary}, a clear and economically coherent hierarchy of criteria emerges.

First, downside- and recovery-related indicators dominate the screening stage across all market regimes. Recovery time (RT) is the most influential criterion, with average weights between 0.21 and 0.23 and a pronounced peak in bear markets, underscoring the central role of resilience and post-drawdown behavior under stressed conditions. Recovery factor (RF) consistently follows as the second most relevant indicator, with weights around 0.18--0.19 and limited dispersion, confirming the stability of longer-horizon drawdown metrics in providing cross-sectional discrimination. Then, a second group of criteria contributes more uniformly across regimes. Specifically, ESG-related measures display stable and meaningful weights: ESG momentum ranges between 0.084 and 0.097, while $\mathrm{ESG\!-\!VOL}^{24}$ and $\mathrm{ESG\!-\!VOL}^{36}$ lie in the 0.09--0.12 interval, with no pronounced regime-specific peaks. This pattern indicates that short- and medium-horizon ESG variability provide consistent discriminative power independently of market conditions. Similarly, the two asymmetric beta ratios retain significant importance, particularly $\mathrm{UDR}^{0.25}$, while exhibiting the highest temporal variability (standard deviations around 0.04--0.05), reflecting the inherently time-varying nature of conditional betas and their sensitivity to shifts in market states. Finally, a third group comprises criteria whose contribution is strongly regime dependent. Normalized mutual information (NMI) attains its lowest relevance in bear markets (0.0082), compared to bull (0.0203) and sideways (0.0214) phases, suggesting that benchmark co-movement becomes less informative precisely when market-wide risk intensifies and cross-sectional dispersion narrows. In contrast, the technical score $S^{\mathrm{tech}}$ becomes substantially more informative during downturns (0.0830), compared to bull (0.0257) and sideways (0.0471) regimes. This behavior indicates that trend asymmetries, volatility-driven signals, and range-trading dynamics contain richer discriminative content under stressed environments.

\begin{table}[t]
    \centering
    \footnotesize
    \caption{Average MEREC weights across market phases, with standard deviations in parentheses.}
    \label{tab:MEREC_summary}
    \begin{tabularx}{0.9\textwidth}{
        l
        >{\centering\arraybackslash}X
        >{\centering\arraybackslash}X
        >{\centering\arraybackslash}X }
        \hline
        \textbf{Criterion} & \textbf{Bull} & \textbf{Bear} & \textbf{Sideways} \\
        \hline
        $\mathrm{MOM}$               & 0.0663 (0.0178) & 0.0616 (0.0152) & 0.0606 (0.0168) \\
        $\mathrm{UDR}^{0.25}$        & 0.1079 (0.0463) & 0.0997 (0.0473) & 0.0904 (0.0365) \\
        $\mathrm{UDR}^{0.50}$        & 0.0583 (0.0214) & 0.0480 (0.0184) & 0.0605 (0.0285) \\
        $\mathrm{RT}$                & 0.2153 (0.0278) & 0.2292 (0.0351) & 0.2098 (0.0230) \\
        $\mathrm{RF}$                & 0.1837 (0.0094) & 0.1874 (0.0070) & 0.1845 (0.0075) \\
        $\mathrm{ESG\!-\!MOM}$       & 0.0949 (0.0133) & 0.0844 (0.0150) & 0.0969 (0.0132) \\
        $\mathrm{ESG\!-\!VOL}^{24}$  & 0.1054 (0.0196) & 0.0900 (0.0170) & 0.1068 (0.0189) \\
        $\mathrm{ESG\!-\!VOL}^{36}$  & 0.1222 (0.0175) & 0.1086 (0.0150) & 0.1221 (0.0186) \\
        $\mathrm{NMI}$               & 0.0203 (0.0061) & 0.0082 (0.0030) & 0.0214 (0.0049) \\
        $S^{\mathrm{tech}}$          & 0.0257 (0.0093) & 0.0830 (0.0141) & 0.0471 (0.0146) \\
        \hline
    \end{tabularx}
\end{table}

Our findings show that MEREC assigns weights in a statistically stable and economically interpretable manner: recovery-related metrics dominate the hierarchy; ESG, momentum, and asymmetric betas provide a consistent intermediate layer of information; and regime-sensitive criteria — especially NMI and $S^{\mathrm{tech}}$ — gain or lose relevance according to market conditions.

\subsubsection{TODIMSort: limiting profiles and regime sensitivity}

The TODIMSort stage converts the weighted multi‑criteria information into an endogenous assignment of assets to short, neutral, and long categories. A key modeling choice concerns the construction of the limiting profiles, which are defined through regime‑conditioned empirical quantiles. In this respect, the three quantile pairs $(p_1,p_2)\in\{(5,50),(10,60),(20,70)\}$ allow us to vary the selectivity of the profiles in a controlled manner that enables a direct assessment of the robustness of the classification mechanism.

\begin{figure}[t]
\centering
\includegraphics[width=0.45\textwidth]{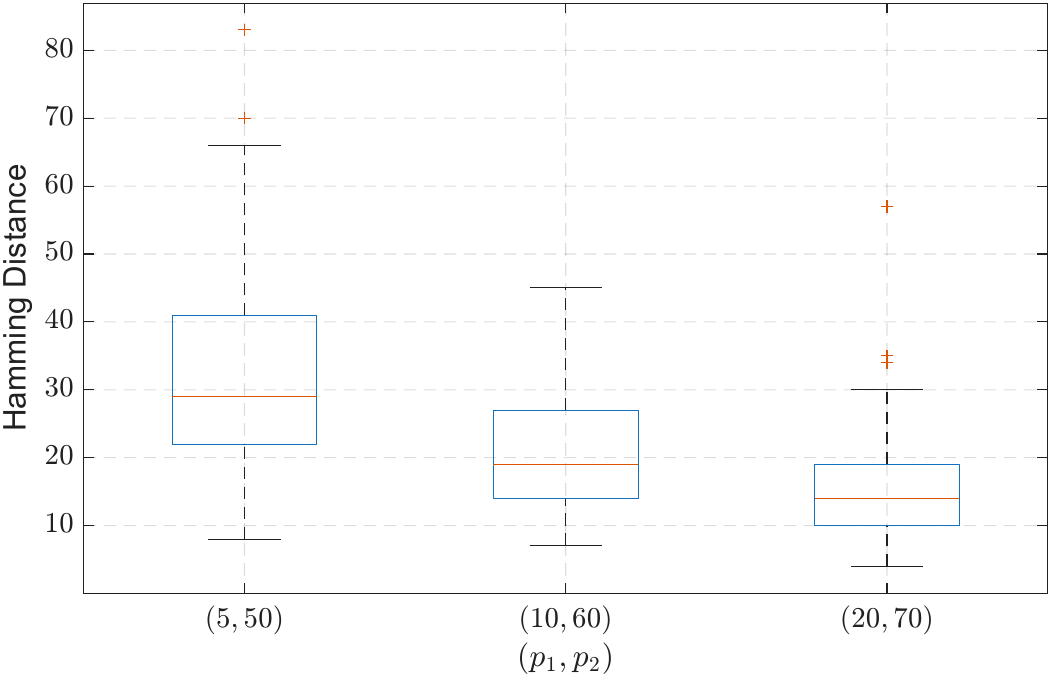}
\includegraphics[width=0.45\textwidth]{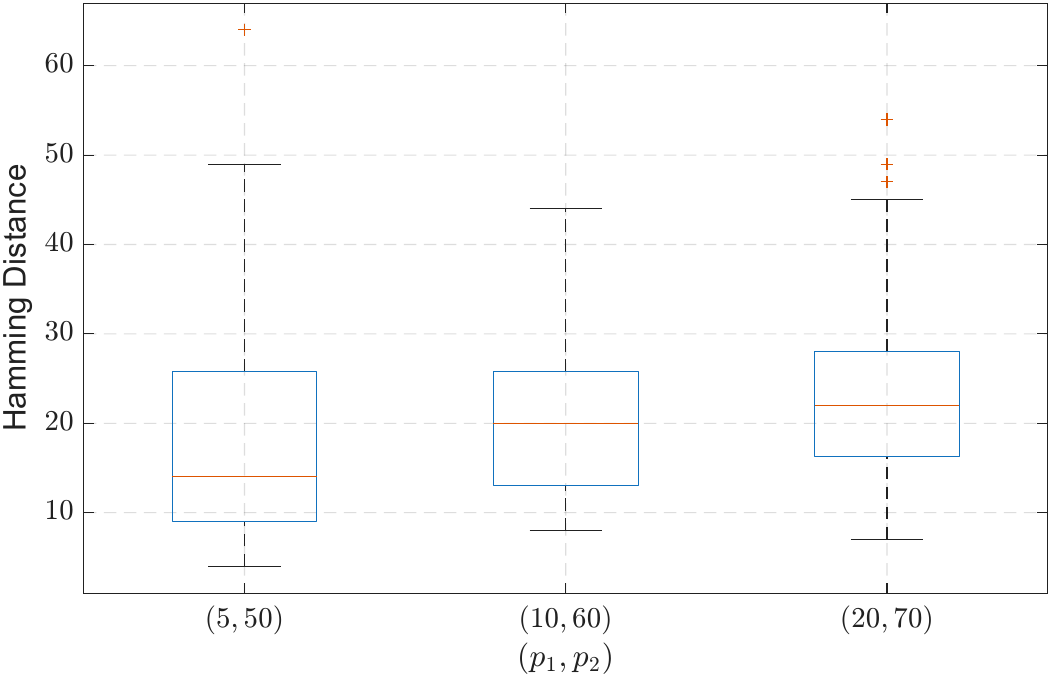}
\caption{Box plot of the Hamming distance for long (left) and short (right) legs under the three limiting-profile configurations.}
\label{fig:boxplotHamming}
\end{figure}

The impact of these alternative specifications on portfolio rotation can be quantified by tracking changes in class membership among consecutive rebalancing dates. To this end, we measure a Hamming‑type distance between the label vectors at $t$ and $t{+}1$. Let $\ell_{t,i}\in\{-1,0,+1\}$ denote the class assigned to asset $i$ at date $t$, and define the full label vector $\boldsymbol{\ell}_t=(\ell_{t,1},\ldots,\ell_{t,N})^\top$.\footnote{Specifically, $\ell_{t,i} = +1$ if $i \in I_{\text{long}}$, it is $-1$ if $i \in I_{\text{short}}$, and 0 otherwise.} The intertemporal Hamming distance is
\begin{equation}
d_{\text{Hamm}}(t,t{+}1)=\sum_{i=1}^{N}\mathbb{I}\!\left(\ell_{t,i}\neq \ell_{t+1,i}\right),
\label{eq:hamming_labels}
\end{equation}
with $\mathbb{I}$ representing the indicator function. This distance counts how many assets change class between $t$ and $t{+}1$. Since our empirical analysis focuses on the rotation of the long and short legs separately, we apply the same definition to the corresponding subsets of indices. Specifically, $d_{\text{Hamm}}^{\text{long}}(t,t{+}1)$ is computed by restricting the sum in Eq.~\eqref{eq:hamming_labels} to the assets that belong to the long leg at either $t$ or $t{+}1$, and analogously for $d_{\text{Hamm}}^{\text{short}}(t,t{+}1)$. This makes explicit that the leg-specific distances shown in Figure~\ref{fig:boxplotHamming} are projections of the general Hamming measure onto the buy-list and sell-list, respectively. In particular, we can see that the three configurations $(5,50)$, $(10,60)$, and $(20,70)$ produce box plots that largely overlap in both panels, indicating that the construction of regime-conditioned profiles is not excessively sensitive to moderate changes in $(p_1,p_2)$. In practice, most of the observed rotation originates from genuine shifts in dominance and from regime changes rather than from arbitrary threshold choices.

Table~\ref{tab:category_counts} reports the average number of assets assigned to the short, neutral, and long categories under each market regime and each $(p_1,p_2)$ setting. Two implications stand out. First, the neutral class consistently contains the majority of the universe (roughly 330--364 assets), indicating that TODIMSort concentrates the tails of the dominance distribution into actionable long and short opportunity sets while preserving a sizable middle region of borderline assets. Second, class sizes respond to regime conditions in an economically coherent way: bear markets yield a larger short set and a smaller long set compared to bull and sideways phases. Moreover, changing the quantile thresholds leads to predictable and contained variations in the sizes of the long and short sets, thus providing a clear way to balance the degree of selectivity against the breadth of the investable universe, without disrupting the stability of the rotation mechanism.

\begin{table}[t]
    \centering
    \footnotesize
    \caption{Average number of assets assigned to long, neutral, and short categories under each market regime (bull, bear, sideways) across the three parameter settings $(5,50)$, $(10,60)$, and $(20,70)$.}
    \label{tab:category_counts}

    \begin{tabularx}{0.95\linewidth}{
        l
        *{9}{>{\centering\arraybackslash}X}
    }
    \toprule
    & \multicolumn{3}{c}{\textbf{(5,50)}} 
    & \multicolumn{3}{c}{\textbf{(10,60)}}
    & \multicolumn{3}{c}{\textbf{(20,70)}} \\
    \cmidrule(lr){2-4} \cmidrule(lr){5-7} \cmidrule(lr){8-10}
    \textbf{Category} 
        & \textbf{Bull} & \textbf{Bear} & \textbf{Side}
        & \textbf{Bull} & \textbf{Bear} & \textbf{Side}
        & \textbf{Bull} & \textbf{Bear} & \textbf{Side} \\
    \midrule
    Short   
        & 32  & 51  & 31 
        & 44  & 37  & 39
        & 52  & 38  & 48 \\
    Neutral 
        & 330 & 349 & 332 
        & 340 & 364 & 345
        & 342 & 358 & 345 \\
    Long    
        & 59  & 20  & 57 
        & 37  & 20  & 38
        & 27  & 25  & 28 \\
    \bottomrule
    \end{tabularx}
\end{table}

Overall, evidences provided by Figures~\ref{fig:phaseIdentifier}, \ref{fig:boxplotMEREC}, \ref{fig:boxplotHamming}, and the summary statistics reported in Tables \ref{tab:MEREC_summary} and \ref{tab:category_counts}, show that the combined MEREC--TODIMSort stage remains coherent with the prevailing market regimes. In addition, it exhibits economically interpretable shifts in criterion relevance, and it is robust to reasonable variations in the quantile parameters of the limiting-profiles, both in terms of rotation stability and in terms of long/short opportunity-set sizes.

\subsection{Evaluation of the AMPSO Algorithm} \label{subsection:algoEvaluation}

We evaluate the algorithmic efficiency of the proposed solver on a benchmark suite constructed to reflect heterogeneous long-short portfolio configurations. We randomly sample without replacement 10 out of the 69 admissible rebalancing dates of the empirical study (the three dates with short-selling ban has not been considered). For each selected date, the MEREC-TODIMSort procedure with $(p_1,p_2) = (5,50)$ for the limiting-profile configurations identified the sets of investable assets. Then, three independent instances of the long–short portfolio optimization problem were generated by varying the leverage parameter in the set $s \in \{0.10, 0.20, 0.30\}$, while setting the upper bound for the weights of assets in the long leg to $w_i^{u}=1+s$ and the lower bound for those in the short leg to $w_{i'}^{\ell}=-s$. This design yields a total of 30 instances of the problem~\ref{eq:omega_opt_long_short}, each characterized by a different market regime, a different composition of the investable universe, and a feasible region that varies with the leverage level. Alternative limiting-profile 
configurations $(10,60)$ and $(20,70)$ exhibit consistent trends with those observed in the above setting, hence we report them in the Supplementary Material.

In the analysis, we compare the proposed AMPSO algorithm with the original PSO‑TVAC, as well as with two PSO‑TVAC variants enhanced through our adaptive operator selection (AOS) strategy: one employing exclusively crossover operators (PSO‑TVAC‑CROSS) and the other relying solely on mutation operators (PSO‑TVAC‑MUT). Table~\ref{Table:paramSettingAlgos} reports the parameter settings used for the PSO‑related and AOS components of the algorithms. The initial population size depends on the number of decision variables $n$ according to the rule $NP = \min\{100, \max \{20,4\log_2 n\}\}$, which follows the recommendations of \cite{Piotrowski2020}. The parameter values for the TVAC component are those suggested in \cite{Deng2012}. The algorithm implementation and the evaluation of the benchmark suite was conducted in MATLAB r2025b on a workstation with an Intel Core i9‑7900X CPU and 16 GB of RAM. The optimization process terminates when the maximum number of function evaluations is reached, i.e. $\mathrm{MAX}_{FES} = 10000 \times n$. To ensure a fair and statistically robust comparison, each solver was executed for 30 independent runs on every problem in the benchmark suite. Random seeds were varied in runs, while all other experimental conditions were kept fixed.

\begin{table}[H] 
    \centering
    \footnotesize
    \caption{Parameter settings for AMPSO and baseline PSO variants.} 
    \label{Table:paramSettingAlgos}
    \begin{tabular}{ll}
    \toprule
    \textbf{Parameter} & \textbf{Value} \\
    \midrule
    Swarm Size & $NP = \min\{100,\, \max\{20,\, 4\log_2 n\}\}$ \\ 
    Max Fitness Eval & $\mathrm{MAX}_{FES} = 10000 \times n$ \\     
    \multirow{2}{*}{TVAC} 
        & $\omega_{\min} = 0.4$, $\omega_{\max} = 0.9$, $c_{1,\min} = c_{2,\min} = 0.5$, \\
        & $c_{1,\max} = c_{2,\max} = 2.5$ \\
    AOS 
        & $n_{\text{par}} = \min\{10, NP\}$, $T_\text{w} = 50$, 
          $p_{\min} = 0.02$, $p_\text{app} = 0.5$ \\
    \bottomrule
    \end{tabular}
\end{table}

\begin{figure}[t]
\centering
\includegraphics[width=0.75\textwidth]{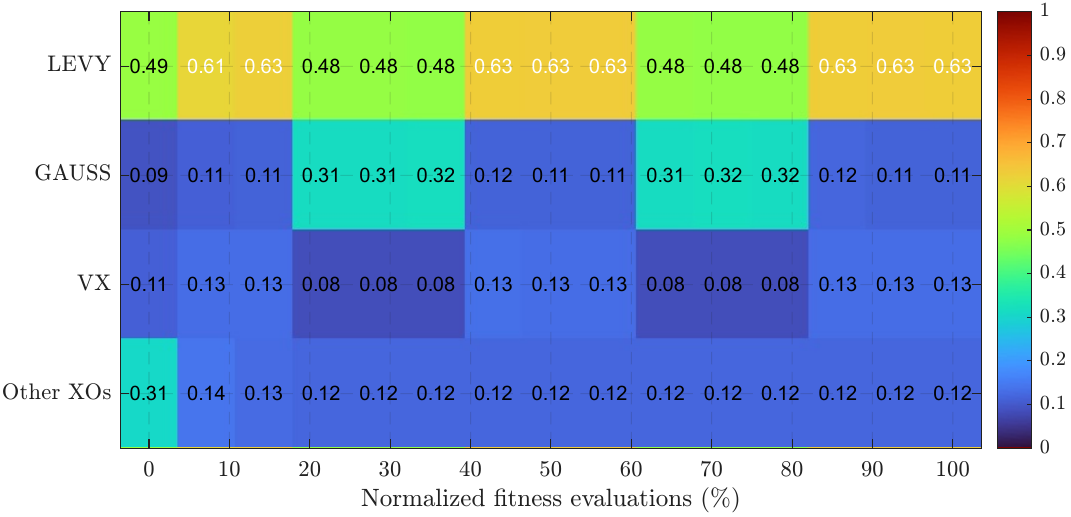}
\caption{Aggregated heatmap of AMPSO operator usage for configuration $(5,50)$ over 15 normalized fitness-evaluation intervals. The first row aggregates the 12 crossover operators into the category “Other XOs”, while the remaining rows report the usage of the vertical crossover (VX), Gaussian mutation (GAUSS), and L\'evy mutation (LEVY). Cell intensities and annotations indicate the proportion of operator applications within each interval.}
\label{fig:heatmapAMPSO}
\end{figure}

Figure~\ref{fig:heatmapAMPSO} provides a synthetic view of the operator‑selection behavior of AMPSO across the entire set of benchmark instances. The heatmap reports the relative frequency with which each operator is selected within 15 equally spaced segments of the total computational budget. A clear pattern emerges. The two mutation operators — Gaussian and Levy — together with the vertical crossover dominate the selection profile throughout most phases of the search. Their consistently high activation rates indicate that AMPSO relies primarily on these mechanisms to balance local refinement and global exploration. In contrast, the remaining crossover operators contribute only marginally. Their activation frequencies remain low and relatively stable across all evaluation intervals, suggesting that they play a secondary role in the dynamics of the algorithm. This behavior is coherent with the structure of the optimization landscape: the combination of mutation‑based perturbations and the vertical crossover appears to provide the most effective means of navigating the feasible region induced by the long–short constraints.

The pairwise comparison among algorithms was first assessed in terms of final performance scores using the Wilcoxon signed ranks test, following \cite{Derrac2011}. Table~\ref{Table:wilcoxon_results} reports, for each competing method, the number of test instances in which AMPSO performs significantly better, equivalent, or worse at the $5\%$ significance level. The comparison with PSO-TVAC is straightforward: AMPSO secures a large majority of significant wins and virtually no losses, clearly outperforming the baseline variant. A similar, though slightly less pronounced, pattern is observed against PSO-TVAC-MUT, where AMPSO still achieves substantially more significant wins than defeats. In contrast, the comparison with PSO-TVAC-CROSS is notably more balanced. The numbers of significant wins and losses are comparable, and most cases fall under statistical equivalence. This indicates that the crossover-only variant remains competitive with AMPSO across the benchmark suite, with neither method exhibiting a decisive statistical advantage at the $5\%$ level.

\begin{table}[t]
\centering
\footnotesize
\caption{Number of instances in which AMPSO performs better, equally, or worse than the reference algorithms according to the Wilcoxon signed-rank test at the 5\% significance level.}
\label{Table:wilcoxon_results}
\begin{tabular}{lccc}
\toprule
\textbf{Algorithm} & \textbf{Better} & \textbf{Equal} & \textbf{Worse} \\
\midrule
PSO-TVAC          & 30 & 0 & 0 \\
PSO-TVAC-CROSS    & 4  & 22 & 4 \\
PSO-TVAC-MUT      & 29 & 1 & 0 \\
\bottomrule
\end{tabular}
\end{table}

To assess differences in convergence behavior across solvers, we employ Page's trend test as in \cite{derrac2014}. For each optimization problem and each algorithm, we record objective function values at 15 equally spaced cut points and compute, at each cut point, the rank of the competing methods. Let $L$ denote the Page statistic obtained by aggregating these ranks with increasing weights that emphasize later stages of the optimization process. We test the ordered one-sided hypothesis
\begin{align*}
& H_0: \text{AMPSO and the competitor have the same convergence pace}
\\
& H_1: \text{AMPSO converges faster than the competitor}   
\end{align*}
which corresponds to smaller objective values in our case. In our case, since the empirical Omega ratio produces non-convex and non-smooth surfaces without guaranteed plateaus, and objective values typically continue to evolve until the evaluation budget is exhausted, we can not apply the ``Alternative'' Page trend test formulation reported in \cite{derrac2014}. We only report the results for the standard form of the Page statistic in Table \ref{Table:Page_results}. In this context, the emerging pattern is fully consistent with the evidence from the Wilcoxon analysis. AMPSO exhibits a clear and systematic convergence advantage over PSO-TVAC, with $p$-values that strongly reject the null hypothesis of equal convergence profiles, indicating that it not only reaches better final objective function values but also progresses more rapidly throughout the optimization process. A similar advantage is observed in comparison with PSO-TVAC-MUT, where the trend test again favors AMPSO in most instances. The mutation-only variant does not match the rate at which AMPSO reduces the objective value, suggesting that adaptive selection of variation operators provides a more effective exploratory–exploitative balance. In contrast, the comparison with PSO-TVAC-CROSS yields more tempered results. In several cases, the $p$-values do not indicate a statistically significant trend in favor of either method, suggesting that the crossover-only variant achieves a convergence profile comparable to that of AMPSO. 

Collectively, the results of the Page trend test confirm that the improvements achieved by AMPSO relative to PSO-TVAC and PSO-TVAC-MUT derive not only from superior end-of-run performance, but also from a consistently faster convergence pattern. However, PSO-TVAC-CROSS remains competitive in convergence speed, showing no systematic disadvantage when the temporal evolution of the optimization process is taken into account.

\begin{table}[t]
\centering
\footnotesize
\caption{Page trend test comparing the convergence speed of AMPSO with baseline solvers. 
The statistic $L$ is computed as the weighted sum of within-problem ranks across the 15 cut points. 
The one-sided $p$--value tests the ordered alternative that AMPSO exhibits faster convergence than the competitor. 
Significant values ($<5\%$) are in bold.}
\label{Table:Page_results}

\begin{tabular}{lccc}
\toprule
\textbf{Comparison} & $L$ & $Z$ & $p$--value \\
\midrule
AMPSO vs PSO--TVAC        & 30375 & 3.8414  & \textbf{6.12e-05} \\
AMPSO vs PSO--TVAC--CROSS & 29220 & 1.0235  & 0.1530 \\
AMPSO vs PSO--TVAC--MUT   & 30060 & 3.0729  & \textbf{0.0011} \\
\bottomrule
\end{tabular}
\end{table}

\subsection{Ex-Post Profitability Analysis}

This subsection evaluates the ex‑post performance of the long–short portfolios produced by the MEREC–TODIMSort–AMPSO framework. Building on the optimization setup introduced in Section~\ref{subsection:algoEvaluation}, we retain the same three leverage levels, $s \in \{0.10, 0.20, 0.30\}$, and focus here on how portfolio quality is affected by the pre‑selection stage. Specifically, we examine three limiting‑profile configurations for TODIMSort, $(p_1,p_2)\in\{(5,50),\, (10,60),\, (20,70)\}$, which span increasing degrees of selectivity in defining the long and short opportunity sets and thus enable a controlled assessment of how stringent classification rules shape portfolio characteristics. Moreover, to assess the contribution of sustainability-related information to long–short portfolio construction, the performance of the ESG-enhanced strategies is compared both with portfolios built under the same parameterization but excluding ESG criteria in the pre-selection phase and with the market-value-weighted index. This design enables us to evaluate whether the incorporation of ESG-based indicators yields systematic improvements in risk-adjusted performance and downside-risk behavior.

\subsubsection{Ex-post performance and risk measures}

Following established principles in performance‑measurement theory, we employ a multidimensional set of ex‑post indicators that capture return efficiency, total risk, downside risk, and persistence of drawdowns \citep{Reilly2018}. Ex‑post performance is evaluated on the monthly sequence of optimized portfolio returns $\{r_{t}^{\text{opt}}\}$ generated in the $M = 72$ investment windows. Let $\bar r$ and $\sigma_{M}$ denote, respectively, the sample monthly mean and volatility of $\{r_{t}^{\text{opt}}\}$. The performance metrics considered are as follows.

\noindent \textbf{Compound annual growth rate.} Let $W_M=\prod_{t=1}^{M}(1+r_{t}^{\text{opt}})$ be the terminal wealth from unit initial capital. The annualized growth rate is
\[
\mathrm{CAGR} = W_M^{\,12/M}-1.
\]

\noindent \textbf{Annualized Sharpe ratio.} This measure represents the compensation earned by the investor per unit of risk, and is computed as
\[
\mathrm{Sharpe}_{\mathrm{ann}}
= \frac{\bar r}{\sigma_M}\, \sqrt{12}.
\]

\noindent \textbf{Annualized Sortino ratio.} To penalize only downside variability, we first define the lower semideviation as 
$\sigma_M^- =
\sqrt{\frac{1}{M}
\sum_{t=1}^{M}\bigl(\min\{r_{t}^{\text{opt}}-\mathrm{MAR},0\}\bigr)^2 }$,
where $MAR$ denotes the minimum acceptable return. Then, the annualized Sortino ratio is
\[
\mathrm{Sortino}_{\mathrm{ann}}
= \frac{\bar r}{\sigma_M^-}\, \sqrt{12}.
\]
Following standard practice in empirical performance analysis, we set $MAR = 0$, which yields a downside-risk assessment relative to the break even level and avoids introducing additional noise from short-term interest-rate fluctuations.

\noindent \textbf{Rachev ratio.} Let $q_{1-\alpha}$ and $q_{\beta}$ be the empirical $(1-\alpha)$ and $\beta$ quantiles of $\{r_{t}^{\text{opt}}\}$, respectively. We define 
expected tail gain (ETG) as the average of portfolio returns that exceed $q_{1-\alpha}$, and expected tail loss (ETL) as the average of returns that fall below $q_{\beta}$. Then, the Rachev ratio is
\[
\mathrm{Rachev}_{\alpha,\beta} = \frac{\mathrm{ETG}}{\mathrm{ETL}}.
\]
While Sharpe and Sortino summarize central tendencies and downside dispersion, the Rachev ratio contrasts extreme gains with extreme losses, thus capturing tail asymmetry. In our implementation, we set tail parameters $\alpha= \beta=0.05$. 

\noindent \textbf{Annualized standard deviation.} Overall volatility is reported as
\[
\sigma_{\mathrm{ann}} = \sigma_M \, \sqrt{12}.
\]

\noindent \textbf{Maximum drawdown.} Let $V_t$ denote the wealth process implied by $\{r_{t}^{\text{opt}}\}$ and 
$P_t = \max_{s\le t} V_s$ its running peak. 
The maximum drawdown is
\[
\mathrm{MDD} = \max_{t} \frac{P_t - V_t}{P_t}.
\]

\noindent \textbf{Ulcer index.} With $D_t = (P_t - V_t)/P_t$ denoting the drawdown relative to the preceding peak, the Ulcer index captures both the depth and the persistence of drawdowns, and is defined as
\[
\mathrm{UI} = 
\sqrt{\frac{1}{M}\sum_{t=1}^{M} D_t^2}.
\]

\noindent \textbf{Average turnover.} Let $w_{t,j}$ be the post-trade weight of asset $j$ at time $t$, and $\widetilde w_{t^-,j}$ the pre-trade weight, then the one-way turnover at time $t$, with $t = 2, \ldots M$, is
\[
\mathrm{Turnover}_t =
\frac{1}{2}\sum_{j=1}^{N} |w_{t,j}-\widetilde w_{t^-,j}|.
\]
To assess trading activity and approximate transaction costs in the ex‑post period, we rely on the average monthly turnover.

\subsubsection{Discussion of Results}

Table~\ref{tab:expost} reports ex-post performance and downside-risk measures of optimized portfolios for the three configurations of the limiting-profiles and leverage levels. A cross-configuration inspection reveals several consistent patterns. First, moderate and higher leverage levels generally translate into improved risk-adjusted performance, as reflected by the Sharpe and Sortino ratios, with configuration $(20,70)$ producing the most stable gains across $s$. Second, downside-risk indicators (maximum drawdown and Ulcer index) remain well controlled and exhibit limited dispersion despite the increased long exposure $1+s$. Finally, portfolio turnover remains moderate and increases only gradually with leverage, suggesting that the combination of MEREC–TODIMSort screening and AMPSO optimization yields stable long–short allocations even when the limiting-profile becomes more selective.

\begin{table}[t]
\centering
\footnotesize
\caption{Ex-post performance and risk measures for the optimized portfolio across the three limiting-profile configurations $(5,50)$, $(10,60)$, and $(20,70)$, evaluated at leverage levels $s=0.10$, $0.20$, and $0.30$.}
\label{tab:expost}
\begin{tabularx}{\textwidth}{
c
c
>{\centering\arraybackslash}X
>{\centering\arraybackslash}X
>{\centering\arraybackslash}X
>{\centering\arraybackslash}X
>{\centering\arraybackslash}X
>{\centering\arraybackslash}X
>{\centering\arraybackslash}X
>{\centering\arraybackslash}X}
\hline
\textbf{Config} & \textbf{$s$} & \textbf{CAGR} & \textbf{Sharpe (ann.)} & \textbf{Sortino (ann.)}
& \textbf{Rachev} & \textbf{Std (ann.)} & \textbf{Max DD} & \textbf{Ulcer} & \textbf{Avg Turnover} \\
\hline

\multirow{3}{*}{(5,50)}
& $0.10$ & 0.1785 & 1.0096 & 1.5521 & 0.8724 & 0.1799 & 0.2336 & 0.0826 & 0.7240 \\
& $0.20$ & 0.1440 & 0.7985 & 1.2031 & 0.8532 & 0.1929 & 0.3830 & 0.1438 & 0.8389 \\
& $0.30$ & 0.1726 & 0.8675 & 1.3417 & 1.0172 & 0.2104 & 0.3813 & 0.1383 & 0.9557 \\
\hline

\multirow{3}{*}{(10,60)}
& $0.10$ & 0.1436 & 0.8372 & 1.2549 & 0.7921 & 0.1810 & 0.2973 & 0.1197 & 0.6670 \\
& $0.20$ & 0.1269 & 0.7583 & 1.1067 & 0.7702 & 0.1800 & 0.3623 & 0.1461 & 0.7568 \\
& $0.30$ & 0.1320 & 0.7828 & 1.1308 & 0.7692 & 0.1804 & 0.3009 & 0.1266 & 0.9288 \\
\hline

\multirow{3}{*}{(20,70)}
& $0.10$ & 0.1438 & 0.8124 & 1.2266 & 0.8818 & 0.1882 & 0.3894 & 0.1532 & 0.6104 \\
& $0.20$ & 0.1698 & 0.9083 & 1.4306 & 0.9741 & 0.1946 & 0.3462 & 0.1286 & 0.7346 \\
& $0.30$ & 0.1860 & 0.9293 & 1.5107 & 1.0878 & 0.2080 & 0.3265 & 0.1087 & 0.8588 \\
\hline

\end{tabularx}
\end{table}

Table~\ref{tab:expost_noESG} shows the corresponding results for portfolios constructed without ESG criteria. Eliminating ESG-based indicators generally leads to a deterioration in both performance and downside-risk behavior. Across most configurations and leverage levels, Sharpe and Sortino ratios decrease, while drawdown measures increase, indicating a reduced ability to mitigate adverse outcomes. Turnover is consistently higher, reflecting less stable long–short allocations when sustainability-related information is excluded.

\begin{table}[t]
\centering
\footnotesize
\caption{Ex-post performance and risk measures for the comparison models corresponding to the limiting-profile configurations $(5,50)$, $(10,60)$, and $(20,70)$, evaluated at leverage levels $s = 0.10$, $0.20$, and $0.30$, together with the benchmark portfolio. In all comparison models, metrics are obtained by excluding ESG criteria from the MEREC-TODIMSort preselection stage.}
\label{tab:expost_noESG}
\begin{tabularx}{\textwidth}{
c
c
>{\centering\arraybackslash}X
>{\centering\arraybackslash}X
>{\centering\arraybackslash}X
>{\centering\arraybackslash}X
>{\centering\arraybackslash}X
>{\centering\arraybackslash}X
>{\centering\arraybackslash}X
>{\centering\arraybackslash}X}
\hline
\textbf{Config} & \textbf{$s$} & \textbf{CAGR} & \textbf{Sharpe (ann.)} & \textbf{Sortino (ann.)}
& \textbf{Rachev} & \textbf{Std (ann.)} & \textbf{Max DD} & \textbf{Ulcer} & \textbf{Avg Turnover} \\
\hline

\multirow{3}{*}{(5,50)}
& $0.10$ & 0.1066 & 0.5886 & 0.8702 & 0.9054 & 0.2106 & 0.3719 & 0.1468 & 0.8182 \\
& $0.20$ & 0.1497 & 0.7904 & 1.2242 & 0.9587 & 0.2037 & 0.3769 & 0.1485 & 0.9384 \\
& $0.30$ & 0.1588 & 0.7737 & 1.2702 & 1.1279 & 0.2235 & 0.3770 & 0.1441 & 1.1613 \\
\hline

\multirow{3}{*}{(10,60)}
& $0.10$ & 0.1722 & 0.9230 & 1.5106 & 1.0478 & 0.1933 & 0.2981 & 0.0978 & 0.7420 \\
& $0.20$ & 0.1079 & 0.6301 & 0.9609 & 0.9333 & 0.1927 & 0.4015 & 0.1463 & 0.8885 \\
& $0.30$ & 0.1325 & 0.6967 & 1.0589 & 0.9561 & 0.2118 & 0.3614 & 0.1321 & 1.0572 \\
\hline

\multirow{3}{*}{(20,70)}
& $0.10$ & 0.1512 & 0.7979 & 1.2159 & 0.9083 & 0.2035 & 0.3729 & 0.1440 & 0.7528 \\
& $0.20$ & 0.1368 & 0.8043 & 1.3161 & 1.1184 & 0.1801 & 0.3301 & 0.1242 & 0.8140 \\
& $0.30$ & 0.1646 & 0.8643 & 1.3734 & 1.0496 & 0.2006 & 0.3259 & 0.1118 & 0.9504 \\
\hline \hline

\textbf{Benchmark} & -- & 0.1964 & 1.0561 & 1.8839 & 1.3669 & 0.1873 & 0.2942 & 0.0965 & -- \\
\hline

\end{tabularx}
\end{table}

Overall, while the benchmark achieves the strongest tail‑sensitive performance, ESG‑enhanced portfolios exhibit systematically lower downside risk and more stable risk‑adjusted returns. These improvements are consistent with the evidence in 
\citet{magnani2024}, who show that firms with more stable ESG profiles — captured through lower ESG‑score volatility — earn higher Sharpe ratios and significant alphas relative to their less stable counterparts. This suggests that ESG‑based filtering can contribute to more resilient long–short allocations by favouring issuers with sustained and predictable ESG trajectories. Finally, a formal statistical assessment of the differences in Sharpe ratios and return variances across ESG, no‑ESG, and benchmark portfolios was also conducted using 
HAC‑robust bootstrap procedures and log‑variance tests. The detailed numerical results, together with the corresponding $p$‑values for the three limiting‑profile 
configurations and leverage levels, are reported in the Supplementary Material. These tests corroborate the qualitative evidence discussed above: ESG‑enhanced portfolios tend to exhibit lower downside risk and more stable risk‑adjusted performance than their no‑ESG counterparts, especially under the baseline configuration $(5,50)$ and moderate leverage. 

\section{Conclusions and Future Works} \label{section:conclusions}

This study demonstrates how a modular combination of multi‑criteria asset classification and adaptive swarm‑based optimization can be effectively used to design ESG‑aware long–short portfolios. At the screening and classification level, the TODIMSort framework, enhanced with MEREC objective weights, assigns assets to preference‑ordered categories in a manner that naturally supports both long‑only and long–short constructions. At the optimization level, the AMPSO solver maximizes the Omega ratio subject to realistic budget, bound, and leverage constraints. To address this non-convex setting, we develop AMPSO, an adaptive particle swarm variant equipped with a multi-operator selection controller and a projection-based repair mechanism for constraint handling. The resulting MEREC–TODIMSort–AMPSO process offers a flexible architecture in which screening, classification, and optimization remain distinct but fully integrable components, allowing each module to be adapted to alternative sets of criteria, realistic constraints, and specific investor objectives.

Empirical analysis of European equity data illustrates the effectiveness of the proposed system. The MEREC–TODIMSort module provides stable long, neutral, and short partitions that adapt to market phases and preserve economic interpretability. The AMPSO solver demonstrates robust convergence behavior and high performance in navigating non-convex and constrained search scenarios. Finally, ex post performance evaluation shows that ESG-aware strategies remain competitive and often outperform their non-ESG counterparts and the market-value-weighted benchmark, particularly in terms of downside protection and drawdown mitigation.

Despite its strengths, the framework presents several limitations that motivate future research. First, the empirical study focuses exclusively on the European market, where ESG data disclosure is relatively advanced. Extending the analysis to regions with different regulatory contexts would offer a broader perspective on the role of sustainability information in heterogeneous markets. Second, turnover limits and explicit transaction cost constraints are not modeled in the current formulation. Although this omission is less restrictive for institutional investors, the integration of such frictions would improve the operational relevance of the framework and further enhance its practical applicability.








\bibliographystyle{elsarticle-harv} 
\bibliography{mainBiblio.bib} 

\end{document}



\begin{center}
{\Large \textbf{Supplementary Material\\
for \\
``A Two-Stage Decision Support System for Sustainability-Aware Long–Short Portfolio Optimization''}}\\[12pt]
\end{center}

\section*{Purpose of this Document}

This Supplementary Material provides additional methodological components and extended numerical analyses that complement the main manuscript. It includes: (i) the complete flow chart of the two-stage MEREC--TODIMSort--AMPSO pipeline; (ii) detailed definitions, taxonomies, and parameter settings for all crossover and mutation operators used in the adaptive operator selection strategy; (iii) operational definitions of the technical indicators and information-theoretic
quantities employed in the screening and regime-detection modules; (iv) operator-usage heatmaps and statistical robustness analyses obtained under the alternative limiting-profile configurations $(10,60)$ and $(20,70)$; (v) the full pseudocode and implementation details of the AMPSO solver; and (vi) the statistical tests used to assess differences in Sharpe ratios and return
variances across ESG, no-ESG, and benchmark portfolios.

\section{Flow Chart of the Two-Stage Procedure}

\begin{sidewaysfigure}
    \centering
    \includegraphics[angle=-90, width=\linewidth]{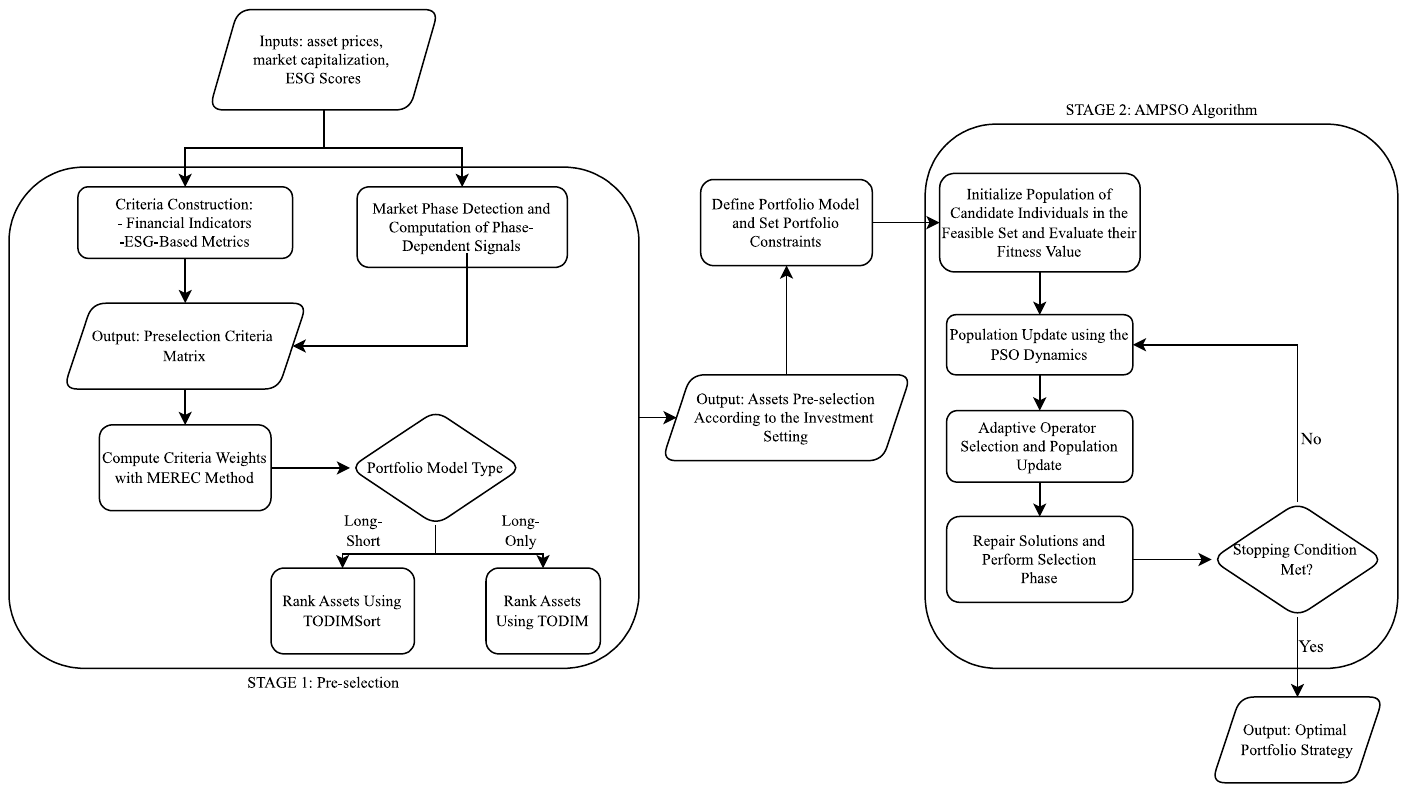}
    \caption{Flow chart summarizing the full two-stage MEREC-TODIMSort-AMPSO procedure, including asset pre-selection and portfolio optimization.}
    \label{Fig:flowchart}
\end{sidewaysfigure}

The complete workflow of the two-stage MEREC–TODIMSort–AMPSO procedure is summarized in the flow chart reported in Figure~\ref{Fig:flowchart}.

\section{Variation Operators}

This section reports the crossover and mutation operators employed in AMPSO. The underlying taxonomy, together with the operator descriptions and parameter settings, follows the classification of \citet{Engelbrecht2016} and the
corresponding original references. All operators act directly on particle positions, with random scalars drawn from the uniform distribution on $[0,1]$ and random vectors sampled component-wise unless otherwise stated.

\subsection*{A. Two-parent real-coded crossover operators}

\paragraph{Arithmetic crossover (AX, \cite{Lovberg2001})}
Given two parents $y_a,y_b\in\mathbb{R}^n$, the offspring is obtained as a convex combination:
\[
y' =
\begin{cases}
r\odot y_a + (1-r)\odot y_b, & \tilde{r} < p_{\mathrm{AX}},\\[4pt]
(1-r)\odot y_a + r\odot y_b, & \tilde{r} \ge p_{\mathrm{AX}}
\end{cases}
\]
where $r\sim U([0,1])^n$ and $\tilde{r}\sim U([0,1])$. The parameter $p_{\mathrm{AX}}\in[0,1]$ controls the probability of selecting one of the two symmetric blending directions. AX preserves feasibility in convex regions and improves the exploitation around parent solutions.

\paragraph{Blend crossover (BLX-$\alpha$, \cite{Duong2010})}
Given a target particle $y_j$ and a parent $y_a$, the offspring components are generated
as
\[
y' = (1-\gamma)\odot y_j + \gamma\odot y_a
\]
with $\gamma = (1+2\alpha)\,U([0,1])^n$ and $\alpha>0$.
BLX-$\alpha$ expands the search interval beyond the parental segment, favoring
exploration.

\paragraph{Horizontal crossover (HX, \cite{Meng2016})}
Given two parents $y_a,y_b$, an offspring is sampled as
\[
y' = r\odot y_a + (1-r)\odot y_b + c\odot (y_a - y_b)
\]
where $r\sim U([0,1])^n$ and $c\sim U([-1,1])^n$.
The perturbation term $c\odot (y_a-y_b)$ allows exploration beyond the convex
hull of the parents.

\subsection*{B. Multi-parent real-coded crossover operators}

\paragraph{Normalized multi-parent crossover (NMPCO, \cite{Wang2008})}
Given four parents $\{y_j,y_a,y_b,y_c\}$, the offspring is defined as
\[
y' = k_1 y_j + k_2 y_a + k_3 y_b + k_4 y_c
\]
where $k_s$, $s = 1,\ldots,4$ are normalized coefficients:
\[
k_s = \frac{\kappa_s}{\sum_{s'=1}^{4} \kappa_{s'}},\qquad
\kappa_s\sim U([0,1]).
\]
This operator exploits multiple search directions, increasing diversity.

\paragraph{Parent-centric crossover (PCX, \cite{Deb2002})}
Let $\{y_1,\ldots,y_{n_\mu}\}$ be sampled parents and
$\bar{y}$ their centroid.
Let $y_a$ be the index parent, with direction $d_a = y_a - \bar{y}$.
The offspring is
\[
y' = y_a
+ N(0,\sigma_1^2)\,\frac{d_a}{\|d_a\|}
+ \sum_{l\neq a} N(0,\sigma_2^2)\,\delta_l\, e_l
\]
where $e_l$ form an orthonormal basis orthogonal to $d_a$,
$\delta_l$ are perpendicular distances, and $\sigma_1,\sigma_2>0$ control spread.

\paragraph{Random-index PCX (PCXr)}
As PCX, but the index parent $y_a$ is selected randomly, improving diversity.

\paragraph{Parent-centric crossover with global guidance (PSPG)}
A variant of PCX incorporating the global best particle as one of the parents,
biasing offspring toward promising regions while maintaining diversity.

\subsection*{C. Discrete crossover operators \citep{Engelbrecht2013}}

\paragraph{One-point personal best crossover (DX-$(y^*,1)$)}
A single index $s_0$ is sampled uniformly, and
\[
y'_s =
\begin{cases}
y^*_{j,s}, & s=s_0,\\
y_{j,s},   & \text{otherwise}.
\end{cases}
\]

\paragraph{Uniform personal best crossover (DX-$(y^*,u)$)}
Each component is inherited from the personal best with probability $p_c$:
\[
y'_s =
\begin{cases}
y^*_{j,s}, & \tilde{r} < p_c,\\
y_{j,s},   & \text{otherwise}.
\end{cases}
\]

Analogous operators DX-$(\hat{y},1)$ and DX-$(\hat{y},u)$ copy loci from the global best $\hat{y}$, while blended-reference variants DX-$(y^*,\hat{y},1)$ and
DX-$(y^*,\hat{y},u)$ exploit a convex combination
$r'y_j^* + (1-r')\hat{y}$, with $r'\sim U([0,1])$.

\paragraph{Vertical crossover (VX, \cite{Meng2016})}
Given a parent $y_a$, two indices $d_1,d_2$ are sampled and component $d_1$ is updated as
\[
y'_{d_1} = \tilde{r}\, y_{a,d_1} + (1-\tilde{r})\, y_{a,d_2}, \qquad \tilde{\tilde{r}}\sim U([0,1])
\]
leaving all other coordinates unchanged.
This intra-parent recombination promotes local diversification.

\subsection*{D. Mutation operators}

\paragraph{Gaussian mutation (GAUSS, \cite{HigashiIba2003})}
Each component undergoes a perturbation
\[
y' = y_i + \sigma\,\epsilon_i,\qquad \epsilon_i\sim N(0,1)
\]
with $\sigma>0$ controlling step size.
This operator supports fine-grained local search.

\paragraph{Levy mutation (LEVY, \cite{Gao2023})}
Occasional long jumps are introduced via
\[
y' = y_i + \alpha'\,\lambda_i,
\]
where $\lambda_i$ is sampled from a Levy distribution and $\alpha'>0$ controls scale. The heavy-tailed behavior helps escape local minima and encourages global exploration.

\subsection*{E. Parameter settings}

Table~\ref{tab:operators_summary} reports the crossover and mutation operators employed in AMPSO, together with their acronyms and the parameter settings used in the experimental analysis.

\begin{table}[t]
\centering
\footnotesize
\caption{Summary of crossover and mutation operators used in AMPSO, including acronyms
and parameter settings adopted in the experiments.}
\label{tab:operators_summary}

\begin{tabular}{lp{1.9cm}p{6.6cm}}
\toprule
\textbf{Operator} & \textbf{Acronym} & \textbf{Parameters / Notes} \\
\midrule
Arithmetic crossover & AX &
$p_{\text{AX}} = 0.5$; $r \sim U([0,1])^n$; convex combination of parents. \\

Blend crossover & BLX-$\alpha$ &
$\alpha = 0.2$; exploration beyond parental segment. \\

Horizontal crossover & HX &
$r \sim U([0,1])^n$, $c \sim U([-1,1])^n$; includes directional perturbation. \\

Normalized multi-parent crossover & NMPCO &
Four parents; coefficients $k_s$ normalized from $\kappa_s \sim U([0,1])$. \\

Parent-centric crossover & PCX &
$\sigma_1 = 0.1$, $\sigma_2 = 0.9$; index parent selected deterministically. \\

Random-index PCX & PCXr &
As PCX, but index parent selected at random. \\

Parent-centric crossover with global\\guidance & PSPG &
As PCX, with global best included among parents. \\

One-point personal best crossover & DX-$(y^\ast,1)$ &
One index $s_0$ sampled uniformly; single-point inheritance. \\

Uniform personal best crossover & DX-$(y^\ast,u)$ &
Inheritance from personal best with probability $p_c = 0.5$. \\

One-point global best crossover & DX-$(\hat{y},1)$ &
As above, using global best locus. \\

Uniform global best crossover & DX-$(\hat{y},u)$ &
As above, with inheritance probability $p_c = 0.5$. \\

Blended reference crossover & DX-$(y^*,\hat{y},1)$, DX-$(y^*,\hat{y},u)$ &
Blending via $r' \sim U([0,1])$:
$y_{\text{mix}} = r' y^\ast + (1 - r')\hat y$. \\

Vertical crossover & VX &
Two indices $d_1,d_2$ sampled;
update $y'_{d_1} = \tilde{\tilde{r}} y_{d_1} + (1-\tilde{\tilde{r}}) y_{d_2}$. \\
\midrule

Gaussian mutation & GAUSS &
$\sigma = 0.05$. \\

Levy mutation & LEVY &
$\alpha' = 1.5$. \\
\bottomrule
\end{tabular}
\end{table}

\section{AMPSO: Full Pseudocode and Implementation Details}

Algorithm \ref{PSEUDOCODE:AMPSO} summarizes the proposed AMPSO procedure. Specifically, lines 1–5 describe the initialization phase, where operator selection probabilities are set uniformly, the swarm is randomly generated within the feasible region, and the objective function is evaluated to initialize personal and global best solutions. Lines 6–7 introduce the main generational loop and the update of the time-varying acceleration coefficients that govern the PSO-TVAC dynamics. In lines 8–13, each particle evolves through the standard PSO update rule, followed by clamping the velocity, projection onto the feasible space, and evaluation of the resulting trial solution. Lines 14–22 implement the adaptive operator selection mechanism: with a prescribed application probability, a subset of $n_{\text{par}}$ particles is sampled to assess the performance of all available operators in terms of solution quality and dispersion, rewards are computed using the compass-based criterion, credits are assigned over a sliding window, and operator selection probabilities are updated through probability matching. The operator selected by this adaptive controller is then applied to generate the new population of offspring, which is repaired if necessary and evaluated alongside the candidates generated by PSO. Finally, lines 23–26 update the population and advance the generation counter until the termination condition is met, at which point the best feasible solution found is returned.

\resizebox{0.95\textwidth}{!}{
\begin{algorithm}[H]
\footnotesize
\DontPrintSemicolon
\SetAlgoLined
\SetKwInOut{Input}{Input}\SetKwInOut{Output}{Output}
\Input{$NP$, $g_\text{max}$, operator pool $\mathcal{O}=\{o_1,\dots,o_{n_\text{op}}\}$, adaptive operators parameters, $p_\text{app}$, $p_\text{min}$, $T_\text{w}$, $n_\text{par}$}
\Output{$\mathrm{y}_\text{gbest}$}
\BlankLine

Set $g = 0$ \;
\textbf{Initialize} operator probabilities $\mathbf{p}(g) = (p_1(g),\dots,p_{n_\text{op}}(g))$ uniformly\;
\textbf{Initialize} the swarm $\mathcal{P}_g = \left\{y_j(g) \in \mathbb{R}^d \colon j = 1,\ldots,NP\right\}$ by sampling candidates within the feasible set and initialize velocities to zero\;
\textbf{Calculate} the objective function $f(y_j(g))$ for each trial solution in $\mathcal{P}_g$ \;
\textbf{Set} personal bests $y_j^*(0)$ and global best $\hat{y}(0)$.\;
\While{$g < g_{MAX}$}{
    \textbf{Update} $w(g)$, $c_1(g)$, and $c_2(g)$ according to Eqns. (5.3)--(5.5) \; 
    \For{$j = 1:NP$}{
        \tcp{PSO-TVAC Update}
        \textbf{Generate} a trial position $y'_j(g)$ using Eqns. (5.1) and (5.2)
        with parameters $w(g)$, $c_1(g)$, and $c_2(g)$\;
        \textbf{Apply} velocity clamping \;
        \textbf{Repair} $y'_j(g)$ by Eqns. (5.7) and (5.8) \; 
        \textbf{Evaluate} objective function value of $y'_j(g)$\;
        \textbf{Update} personal best $y_j^*(g)$ and global best $\hat{y}(g)$\;
        \tcp{Adaptive Operator Selection}
        \textbf{Draw} $u \sim U(0,1)$ \;
        \If{$u \le p_{app}$}{
            \textbf{Sample} a subset of $n_{par}$ particles and apply each operator $\mathcal{o}_k \in \mathcal{O}$, recording the mean objective value and the mean dispersion \;
            \textbf{Aggregate} improvements over a sliding window and evaluate rewards using the compass projection with angle $\phi_g$ calculated as in (5.6) \; 
            \textbf{Assign} operator credits and obtain $c_k(g)$ \;
            \textbf{Update} operator selection probabilities via probability matching \;
            \textbf{Generate} $y''_j(g)$ using the selected operator\;
            \textbf{Repair} $y''_j(g)$ by Eqns. (5.7) and (5.8) \; 
        }
        \textbf{Update} $y_j^*(g)$ and $\hat{y}(g)$ considering also $y''_j(g)$ particles\;
    }
    \textbf{Set} $\mathcal{P}_{g+1} = \left\{y_j(g+1) \colon j = 1,\ldots,NP\right\}$ \;
    \textbf{Set} $g = g + 1$ \;
}

\textbf{Denote} the solution with the lowest $f$-value as $\mathrm{y}_\text{gbest}$ \;
\caption{Pseudocode of AMPSO (for constrained minimization)} \label{PSEUDOCODE:AMPSO}
\end{algorithm}
}

\section{Technical Indicators}

This section provides concise operational definitions of the technical indicators used in screening and regime detection. For a detailed discussion of their construction
and interpretation, we refer the reader to the standard textbooks in technical analysis by \citet{Murphy1999}, \citet{Achelis2000}, \citet{Bollinger2002}, \citet{KirkpatrickDahlquist2010}, and \citet{Kaufman2013}.

\paragraph{Simple Moving Average (SMA)}
For a lookback window of length $H$, the moving average of the price series
$\{p_{i,t}\}$ is
\[
\mathrm{SMA}\left(\{p_{i,t}\}, H\right)
=\frac{1}{H}\sum_{h=0}^{H-1}p_{i,t-h}.
\]
The SMA smooths short-term price fluctuations and highlights the underlying trend over the selected horizon. Prices persistently above (below) the SMA typically indicate sustained upward (downward) pressure.

\paragraph{Exponential Moving Average (EMA)}
For smoothing parameter $\alpha\in(0,1)$, the EMA is defined recursively as
\[
\mathrm{EMA}_{i,t}
=\alpha\,p_{i,t}+(1-\alpha)\,\mathrm{EMA}_{i,t-1},
\]
with initialization set to the sample mean over the first $H$ observations. The EMA assigns greater weight to recent observations, making it more responsive to new information than the SMA. An increasing EMA reflects short‑term strengthening of the price dynamics.

\paragraph{Relative Strength Index (RSI)}
Let $U_t$ and $D_t$ denote the average positive and negative price changes over the
past $H$ days. The RSI is
\[
\mathrm{RSI}_{i,t}
=100\left(1-\frac{1}{1+U_t/D_t}\right)\in[0,100].
\]
The RSI measures the balance between recent gains and losses. Extreme values capture overbought or oversold conditions, which are informative for identifying potential short-term reversals. In our implementation, we adopt the standard choice $H = 14$ days.

\paragraph{Moving Average Convergence Divergence (MACD)}
Let $\mathrm{EMA}^{(a)}_{i,t}$ and $\mathrm{EMA}^{(b)}_{i,t}$ be EMAs with
short and long horizons $a<b$. The MACD line is
\[
\mathrm{MACD}_{i,t}
=\mathrm{EMA}^{(a)}_{i,t}-\mathrm{EMA}^{(b)}_{i,t}.
\]
The MACD quantifies the momentum of the price trend by measuring the distance between a short-term and a long-term EMA. Positive (negative) values indicate
upward (downward) momentum and help confirm the directional strength of recent movements. In our implementation, we use the standard settings $a = 12$ and
$b = 26$ days for the short- and long-term EMAs, respectively.

\paragraph{Average Directional Index (ADX)}
Let $DI_t^{+}$ and $DI_t^{-}$ be the smoothed positive and negative directional indicators. The directional index is
\[
DX_{i,t}=100\,\frac{\lvert DI_t^{+}-DI_t^{-}\rvert}{DI_t^{+}+DI_t^{-}}.
\]
The ADX is obtained by smoothing $DX_{i,t}$ and quantifies trend strength independently of direction: high values indicate persistent directional momentum, whereas low values reflect range-bound conditions. In our
implementation, $DX_{i,t}$ is smoothed using a simple moving average with a window length of $14$ days.

\paragraph{Commodity Channel Index (CCI)}
For each asset $i$, let $p_{i,t}$ denote its closing price at time $t$ and let $h_{i,t}$ and $l_{i,t}$ denote, respectively, the corresponding daily high and daily low prices. The typical price is defined as
\[
T_t=\frac{p_{i,t}+h_{i,t}+l_{i,t}}{3}.
\]
Let $\overline{T}_t$ be the $H$–day simple moving average of $\{T_t\}$ and let $\mathrm{MAD}_t$ denote the associated mean absolute deviation,
\[
\mathrm{MAD}_t=\frac{1}{H}\sum_{h=0}^{H-1}\bigl|T_{t-h}-\overline{T}_t\bigr|.
\]
The Commodity Channel Index is then given by
\[
\mathrm{CCI}_{i,t}
=\frac{T_t-\overline{T}_t}{0.015\,\mathrm{MAD}_t}.
\]
Values below $-100$ typically indicate oversold conditions; values above $+100$ indicate overbought conditions. In our implementation, we set the smoothing window to $H = 20$ days.

\paragraph{Bollinger Bands}
Let $\mu_{i,t}$ and $\sigma_{i,t}$ be the $H$–day moving average and standard deviation.
The bands are
\[
\mathrm{UB}_{i,t}=\mu_{i,t}+k\,\sigma_{i,t},\qquad
\mathrm{LB}_{i,t}=\mu_{i,t}-k\,\sigma_{i,t},
\]
with $H = 20$ and $k=2$ in our implementation. Bollinger Bands compare the current price to its recent volatility. Price excursions outside the bands indicate volatility expansions and can signal the onset of breakouts or, depending on the regime, the likelihood of a return toward the mean.

\paragraph{Stochastic Oscillator ($K$–$D$)}
Let $H_t$ and $L_t$ denote the highest and lowest prices observed over the previous $H$ days. The Stochastic Oscillator is defined as
\[
K_{i,t}
= 100\,\frac{p_{i,t}-L_t}{H_t-L_t},
\qquad
D_{i,t}
= \mathrm{SMA}(\{K_{i,\cdot}\}, H')
= \frac{1}{H'}\sum_{j=0}^{H'-1} K_{i,t-j}.
\]
The indicator locates the current price within its recent high-low range, and crossovers between the $K$ and $D$ lines are often interpreted as early signals of
short-term turning points, particularly in sideways markets. In our implementation, we adopt the standard settings $H = 14$ days and $H' = 3$ days for the smoothing of $D_{i,t}$.

\section{Entropy and Mutual Information}

We summarize here the information–theoretic quantities employed in the computation
of the Normalized Mutual Information (NMI) criterion. For a comprehensive treatment
of entropy, mutual information, and related concepts, we refer the reader to the book by \citet{MacKay2003}.

\paragraph{Shannon entropy}
For a discrete random variable $X$ with probability mass function $p(x)$, the Shannon entropy is
\[
H^{\text{Sh}}(X)=-\sum_{x} p(x)\log p(x).
\]
For a bivariate variable $(X,Y)$ with joint distribution $p(x,y)$,
\[
H^{\text{Sh}}(X,Y)=-\sum_{x}\sum_{y} p(x,y)\log p(x,y).
\]

\paragraph{Mutual Information (MI)}
The mutual information between $X$ and $Y$ is
\[
MI(X;Y)
=\sum_{x}\sum_{y} p(x,y)\log\frac{p(x,y)}{p(x)p(y)}
=H^{\text{Sh}}(X)+H^{\text{Sh}}(Y)-H^{\text{Sh}}(X,Y).
\]
MI measures the strength of statistical dependence between the variables, capturing both linear and non-linear relationships.

\section{Evaluation of the AMPSO Algorithm with Additional Configurations}

\paragraph{Operator-usage heatmaps under alternative limiting-profile settings}

\begin{figure}[t]
\centering
\includegraphics[width=0.75\textwidth]{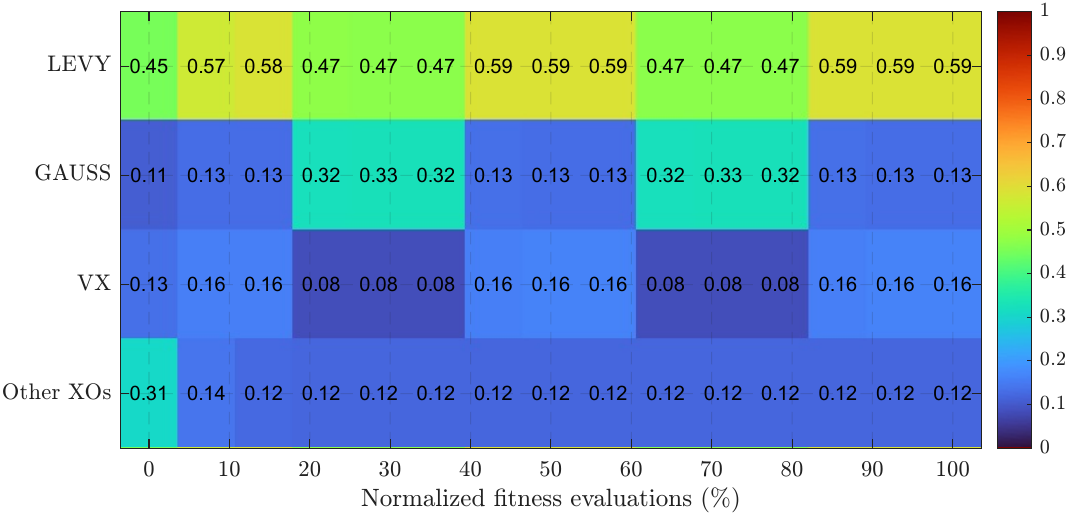}
\caption{Aggregated heatmap of AMPSO operator usage for configuration $(10,60)$ over 15 normalized fitness-evaluation intervals. The first row aggregates the 12 crossover operators into the category “Other XOs”, while the remaining rows report the usage of the vertical crossover (VX), Gaussian mutation (GAUSS), and Levy mutation (LEVY). Cell intensities and annotations indicate the proportion of operator applications within each interval.}
\label{fig:heatmapAMPSO_10_60}
\end{figure}

\begin{figure}[t]
\centering
\includegraphics[width=0.75\textwidth]{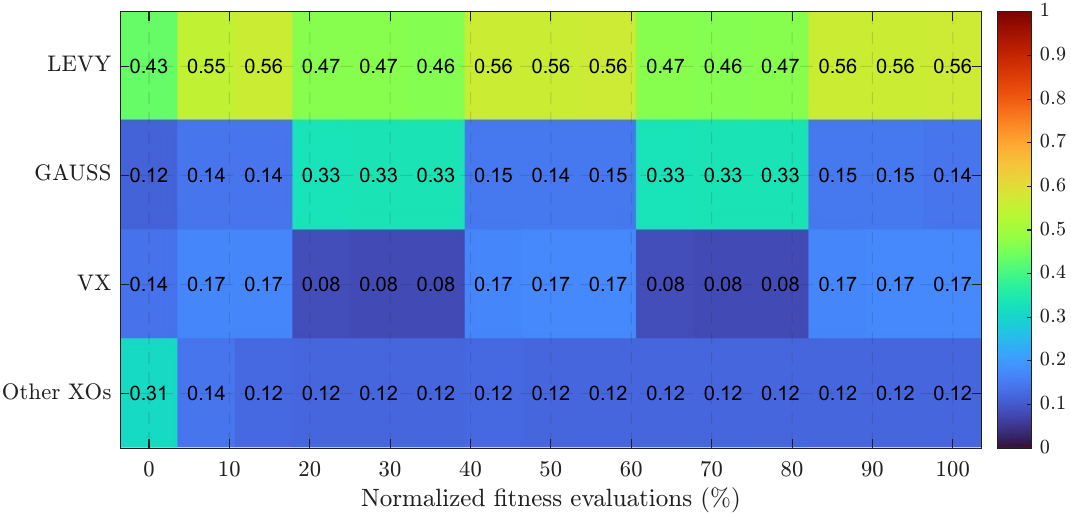}
\caption{Aggregated heatmap of AMPSO operator usage for configuration $(20,70)$ over 15 normalized fitness-evaluation intervals. The first row aggregates the 12 crossover operators into the category “Other XOs”, while the remaining rows report the usage of the vertical crossover (VX), Gaussian mutation (GAUSS), and Levy mutation (LEVY). Cell intensities and annotations indicate the proportion of operator applications within each interval.}
\label{fig:heatmapAMPSO_20_70}
\end{figure}

The operator-selection patterns observed in the configurations $(10,60)$ and $(20,70)$ (Figures~\ref{fig:heatmapAMPSO_10_60} and
\ref{fig:heatmapAMPSO_20_70}) remain aligned with those of $(5,50)$ in the main manuscript. In all settings, AMPSO consistently allocates most of the operator usage to Gauss and Levy mutations, with the vertical crossover
showing a stable intermediate contribution. Only minor numerical differences arise across parameter settings, confirming that the operator-adaptation mechanism is robust to moderate variations in the limiting-profile parameters.

\paragraph{Wilcoxon signed–rank test (5\% level), configurations $(10,60)$ and $(20,70)$}

The pairwise comparisons (Table~\ref{TableS:wilcoxon_results}) corroborate the
evidence reported in the main manuscript for $(5,50)$. AMPSO records a systematic and statistically reliable advantage over PSO-TVAC and PSO-TVAC-MUT in both
additional settings (with at most one tie over 30 instances). In contrast, the comparison with PSO-TVAC-CROSS remains largely balanced: under $(10,60)$ and $(20,70)$ the number of ties increases to 27 out of 30 cases, with only a few significant wins or losses. Overall, the Wilcoxon outcomes confirm that AMPSO's
superiority with respect to PSO-TVAC and PSO-TVAC-MUT is robust to the choice of limiting-profile parameters, whereas PSO-TVAC-CROSS remains statistically comparable to AMPSO on the majority of test instances.

\begin{table}[t]
\centering
\footnotesize
\caption{Wilcoxon signed-rank outcomes (5\% significance level) comparing AMPSO against baseline algorithms for configurations $(10,60)$ and $(20,70)$.}
\label{TableS:wilcoxon_results}

\begin{tabular}{lccc}
\toprule
\multicolumn{4}{l}{\textbf{Configuration $(10,60)$}} \\
\midrule
\textbf{Algorithm} & \textbf{Better} & \textbf{Equal} & \textbf{Worse} \\
\midrule
PSO-TVAC        & 30 & 0  & 0 \\
PSO-TVAC-CROSS & 1  & 27 & 2 \\
PSO-TVAC-MUT   & 30 & 0  & 0 \\
\midrule
\multicolumn{4}{l}{\textbf{Configuration $(20,70)$}} \\
\midrule
\textbf{Algorithm} & \textbf{Better} & \textbf{Equal} & \textbf{Worse} \\
\midrule
PSO-TVAC        & 29 & 1  & 0 \\
PSO-TVAC-CROSS & 2  & 27 & 1 \\
PSO-TVAC-MUT   & 29 & 1  & 0 \\
\bottomrule
\end{tabular}
\end{table}

\paragraph{Page trend test under configurations $(10,60)$ and $(20,70)$}

The Page trend results (Table~\ref{TableS:page_results}) obtained for the additional limiting-profile settings confirm the convergence patterns reported in the main manuscript. In both configurations, AMPSO shows statistically significant faster convergence relative to PSO-TVAC and PSO-TVAC-MUT, while no systematic trend advantage emerges in the comparison with PSO-TVAC-CROSS. These findings closely mirror those obtained under $(5,50)$, indicating that AMPSO's convergence behavior is robust to changes in the limiting-profile parameters.

\begin{table}[t]
\centering
\footnotesize
\caption{Page trend test comparing the convergence speed of AMPSO with PSO-TVAC,
PSO-TVAC-CROSS, and PSO-TVAC-MUT under the limiting-profile configurations $(10,60)$
and $(20,70)$. For each comparison, the Page statistic $L$ aggregates the within-problem
ranks over the 15 cut points, with increasing weights assigned to later evaluations.
The one-sided $p$--value tests the ordered alternative that AMPSO converges faster than
the competitor. Significant values ($<5\%$) are in bold.}
\label{TableS:page_results}

\begin{tabular}{lccc}
\toprule
\multicolumn{4}{l}{\textbf{Configuration $(10,60)$}} \\
\midrule
\textbf{Comparison} & \textbf{$L$} & \textbf{$Z$} & \textbf{$p$-value} \\
\midrule
AMPSO vs PSO--TVAC        & 16200 & 4.1900  & \textbf{1.3947e-05} \\
AMPSO vs PSO-TVAC-CROSS & 15408 & 0.8363 & 0.2015 \\
AMPSO vs PSO-TVAC-MUT   & 16068 & 3.6311  & \textbf{1.4113e-04} \\
\midrule
\multicolumn{4}{l}{\textbf{Configuration $(20,70)$}} \\
\midrule
\textbf{Comparison} & \textbf{$L$} & \textbf{$Z$} & \textbf{$p$-value} \\
\midrule
AMPSO vs PSO--TVAC        & 16200 & 4.1900  & \textbf{1.3947e-05} \\
AMPSO vs PSO-TVAC-CROSS & 15540 & 1.3953  & 0.0815 \\
AMPSO vs PSO-TVAC-MUT   & 15870 & 2.7926  & \textbf{0.002614} \\
\bottomrule
\end{tabular}
\end{table}

\section{Statistical Tests for Performance Differences}

This section reports the statistical tests used to evaluate the significance of the ex-post performance differences discussed in Section~6.4.2 of the main manuscript. We examine differences in Sharpe ratios and return variances across ESG, no-ESG, and benchmark portfolios under the three limiting-profile configurations
$(5,50)$, $(10,60)$, and $(20,70)$ and the leverage levels $s \in \{0.10,0.20,0.30\}$.
The detailed numerical results are provided in Tables~\ref{tab:Sharpe_tests_panels}
and \ref{tab:variance_tests_panels}.

\subsection*{Sharpe Ratio Differences}

Table~\ref{tab:Sharpe_tests_panels} reports the results of pairwise comparisons of annualized Sharpe ratios between ESG, no-ESG, and benchmark portfolios under the
three limiting-profile configurations and leverage settings. To assess whether the Sharpe ratio of one strategy differs significantly from that of another, we adopt the HAC-based bootstrap procedure of \citet{ArdiaBoudt2018}, with $4000$ bootstrap
replications and optimal block length selection. The empirical implementation relies on the \texttt{PeerPerformance} package in R.\footnote{\url{https://cran.r-project.org/web/packages/PeerPerformance/index.html}}
Sharpe ratio differences are evaluated against a one-sided alternative hypothesis consistent with the sign of the observed difference.

Consistent with the findings summarized in the main text, significant differences arise primarily under the baseline configuration $(5,50)$ and moderate leverage ($s = 0.10$), where the ESG strategy outperforms its no-ESG counterpart, while the latter significantly underperforms the benchmark. For the remaining configurations and leverage levels, differences are generally not statistically significant, although the signs of the estimates often favor ESG-based portfolios.

\begin{table}
\centering
\caption{Pairwise comparisons in terms of Sharpe ratio differences between ESG, no-ESG, and benchmark portfolios under three limiting-profile configurations and leverage levels. The Sharpe test follows the approach of \citet{ArdiaBoudt2018} and is based on a HAC bootstrap procedure with 4000 replications and optimal block length selection. The table reports the observed differences between the Sharpe ratios, with $p$-values in parentheses. The alternative hypothesis is one-sided and consistent with the sign of the reported difference, and $p$-values below $5\%$ are highlighted in bold.}
\label{tab:Sharpe_tests_panels}
\small
\begin{tabular}{lccc}
\toprule
\multicolumn{4}{l}{\textbf{Panel A: Configuration (5,50)}}\\
\midrule
 & $s=0.10$ & $s=0.20$ & $s=0.30$ \\
\midrule
ESG vs Benchmark    & $-0.0134 \ (0.8622)$ & $-0.0744 \ (0.3464)$ & $-0.0544 \ (0.4963)$ \\
ESG vs no-ESG       & $0.1215 \ (\textbf{0.0103})$ & $0.0023 \ (0.9917)$ & $0.0271 \ (0.7658)$ \\
no-ESG vs Benchmark & $-0.1349 \ (\textbf{0.0448})$ & $-0.0767 \ (0.2534)$ & $-0.0815 \ (0.5963)$ \\
\midrule
\multicolumn{4}{l}{\textbf{Panel B: Configuration (10,60)}}\\
\midrule
 & $s=0.10$ & $s=0.20$ & $s=0.30$ \\
\midrule
ESG vs Benchmark    & $-0.0631 \ (0.4553)$ & $-0.0859 \ (0.2784)$ & $-0.0789 \ (0.3659)$ \\
ESG vs no-ESG       & $-0.0247 \ (0.6523)$ & $0.0371 \ (0.5843)$ & $0.0249\ (0.7378)$ \\
no-ESG vs Benchmark & $-0.0384 \ (0.5963)$ & $-0.1229\ (0.0729)$ & $-0.1038 \ (0.1919)$ \\
\midrule
\multicolumn{4}{l}{\textbf{Panel C: Configuration (20,70)}}\\
\midrule
 & $s=0.10$ & $s=0.20$ & $s=0.30$ \\
\midrule
ESG vs Benchmark    & $-0.0703 \ (0.3454)$ & $-0.0426 \ (0.6028)$ & $-0.0366\  (0.6768)$\\
ESG vs no-ESG       & $0.0041 \ (0.9467)$ & $0.0301 \ (0.7153)$ & $0.0187 \ (0.7588)$ \\
no-ESG vs Benchmark & $-0.0745 \ (0.2709)$ & $-0.0727 \ (0.3339)$ & $-0.0553 \ (0.4638)$ \\
\bottomrule
\end{tabular}
\end{table}

\subsection*{Variance Differences}

Table~\ref{tab:variance_tests_panels} summarizes the pairwise comparisons in terms of variance differences. To test whether the variance of one portfolio differs from that of another, we use the HAC-robust procedure of \citet{LedoitWolf2011}, which is based on the log-ratio of variances and accounts for both serial dependence and heteroskedasticity. The empirical implementation relies on the MATLAB routines provided by the authors.\footnote{\url{https://www.econ.uzh.ch/en/people/faculty/wolf/publications.html}}
Variance differences are evaluated against a two-sided alternative, with the sign of the estimated difference indicating whether the first portfolio exhibits higher or lower volatility.

Significant differences arise mainly under configuration $(5,50)$ and low leverage ($s = 0.10$), where ESG portfolios exhibit lower variance than their no-ESG
counterparts. For the remaining settings, differences are not statistically significant.

\begin{table}
\centering
\caption{Pairwise comparisons in terms of variance differences between ESG, no-ESG, and benchmark portfolios under three limiting-profile configurations and leverage levels. Variance differences are tested using the HAC-robust procedure of \citet{LedoitWolf2011}. The table reports the estimated differences in log-variances, with $p$-values in parentheses. The alternative hypothesis is two-sided, and $p$-values below $5\%$ are highlighted in bold.}
\label{tab:variance_tests_panels}
\small
\begin{tabular}{lccc}
\toprule
\multicolumn{4}{l}{\textbf{Panel A: Configuration (5,50)}}\\
\midrule
 & $s=0.10$ & $s=0.20$ & $s=0.30$ \\
\midrule
ESG vs Benchmark
& $-0.0002\ (0.6473)$
& $0.0002\ (0.7081)$
& $0.0008\ (0.1924)$ \\

ESG vs no-ESG
& $-0.0010\ (\mathbf{0.0356})$
& $-0.0004\ (0.2885)$
& $-0.0005\ (0.3827)$ \\

no-ESG vs Benchmark
& $0.0008\ (0.1694)$
& $0.0005\ (0.2865)$
& $0.0012\ (\mathbf{0.0256})$ \\

\midrule
\multicolumn{4}{l}{\textbf{Panel B: Configuration (10,60)}}\\
\midrule
 & $s=0.10$ & $s=0.20$ & $s=0.30$ \\
\midrule
ESG vs Benchmark
& $-0.0002\ (0.7513)$
& $-0.0002\ (0.6787)$
& $-0.0002\ (0.6961)$ \\

ESG vs no-ESG
& $-0.0004\ (0.3259)$
& $-0.0004\ (0.3787)$
& $-0.0010\ (\mathbf{0.0500})$ \\

no-ESG vs Benchmark
& $0.0002\ (0.6993)$
& $0.0002\ (0.6695)$
& $0.0008\ (0.3119)$ \\

\midrule
\multicolumn{4}{l}{\textbf{Panel C: Configuration (20,70)}}\\
\midrule
 & $s=0.10$ & $s=0.20$ & $s=0.30$ \\
\midrule
ESG vs Benchmark
& $0.0000\ (0.9504)$
& $0.0002\ (0.6121)$
& $0.0007\ (0.2572)$ \\

ESG vs no-ESG
& $-0.0005\ (0.3277)$
& $0.0005\ (0.2755)$
& $0.0003\ (0.6813)$ \\

no-ESG vs Benchmark
& $0.0005\ (0.3873)$
& $-0.0002\ (0.6525)$
& $0.0004\ (0.3647)$ \\

\bottomrule
\end{tabular}
\end{table}

\bibliographystyle{elsarticle-harv}
\bibliography{SupplementaryMaterialBiblio.bib}